\documentclass[12pt,a4paper]{article}
\usepackage{bbm}
\usepackage{graphicx}
\usepackage{psfrag}
\begin{document}

\begin{flushright}
BI-TP 2001/22 \\
LPT Orsay 01/89
\end{flushright}

\begin{center}  
\large{\bf Dirac operator and Ising model \\ 
on a compact 2D random lattice}
\end{center}

\begin{center}  
L. Bogacz$^{1,2}$, Z. Burda$^{1,2}$, J. Jurkiewicz$^2$, 

A. Krzywicki$^{3}$, C. Petersen$^{1}$ and B. Petersson$^{1}$ 
\end{center}

\centerline{$^{1}$Fakult\"at f\"ur Physik, Universit\"at Bielefeld} 
\centerline{P.O.Box 100131, D-33501 Bielefeld, Germany}
\vspace{0.3cm}
\centerline{$^{2}$Institute of Physics, Jagellonian University} 
\centerline{ul. Reymonta 4, 30-059 Krakow, Poland}
\vspace{0.3cm}
\centerline{$^{3}$ Laboratoire de Physique Th\'eorique, B\^atiment 210,}
\centerline{Universit\'e Paris-Sud, 91405~Orsay, France}

\begin{abstract} \normalsize \noindent
Lattice formulation of a fermionic field theory defined 
on a randomly triangulated compact manifold is discussed,
with emphasis on the topological problem of defining spin structures
on the manifold. An explicit construction 
is presented for the two-dimensional case and its relation
with the Ising model is discussed. Furthermore, an 
exact realization of the Kramers-Wannier duality for the
two-dimensional Ising model on the manifold is 
considered. The global properties of the field are
discussed. The importance of the GSO projection is
stressed. This projection has to be performed for
the duality to hold.

\end{abstract}

\section*{Introduction}
The massless Majorana free fermion theory
belongs to the same universality class as the critical Ising model
on a regular lattice \cite{o,lms,p,pop}. An explicit
construction of the Majorana-Dirac-Wilson fermion field theory 
on a randomly triangulated plane was introduced in \cite{bm}.
This theory was shown to be equivalent to the Ising model also
outside the critical region.
In ref.\cite{bm} Cartesian coordinates were assigned to the nodes  
of the lattice. The directions of the links and of the related 
gamma matrices were expressed in the global frame of the plane. 
This approach works for lattices embedded in a flat background where 
one has at one's disposal a global frame of the underlying 
geometry \cite{cfl1,cfl2,cfl3}. However, if one wants to 
generalize it to a lattice on a curved background where 
no global frame exists, a field of local frames \cite{r,bjk,bjk1} has 
to be introduced. This being done, one can put fermions on a 
curved manifold with any topology and one can eventually attack, for
example, problems of field theory on a dynamical 
geometry like those encountered in string theory or in 
quantum gravity \cite{d1,kkm,d2,bbkp,bbkptt}.

This generalization was partially carried out in \cite{bjk,bjk1} 
where an explicit construction of the Majorana-Dirac-Wilson
operators on curved compact two-dimensional lattices was introduced.

Here we extend these studies. In particular, we discuss 
the significance of the GSO projection, which as in string 
theory also here plays an important physical role \cite{gso, gsw}.
We show that with a careful treatment of the global 
properties of the Dirac operator and of the spin structures 
on the manifold one can find a strict mathematical
one-to-one equivalence between the partition function 
of the Majorana-Wilson fermions and that 
of the Ising model. We show explicitly that 
in our discretization of the Dirac operator on a compact 
manifold, the GSO projection - the summation over all 
spin structures - does remove the non-contractible 
fermionic loops, that is those not corresponding to 
the domain-walls of the corresponding Ising model. 

Further, we show that for the duality to hold exactly
as a one-to-one map between the Ising model on a 
triangulation and on its dual lattice, a sort of 
GSO projection has also to be done. Different spin 
structures for the Ising field are simulated
by physical cuts produced by the introduction 
of antiferromagnetic loops, which mimic antiperiodic 
fermionic boundary conditions.

The paper is organized as follows. In section \ref{secP} 
we give an introduction to the problem of defining the 
Dirac operator on a compact manifold. It is text-book 
material \cite{gsw,top1}. We recall it here for completeness, 
to keep the article self-contained. In section \ref{secDS}, 
we show how to adapt the standard Wilson discretization scheme 
of fermions on the regular translationally invariant hypercubic 
lattice \cite{w} to the local-frame description, which can
be generalized to the case of irregular curved lattices. 
In section \ref{secTP}, using as an example the standard 
toroidal regular lattice, we discuss the sign problem and
the global properties of the fermionic field 
on a compact manifold. In section \ref{secLF} 
we argue that in the case of irregular lattices 
the local frame description is particularly natural, 
and then in section \ref{secSR} we show
how to lift this construction to the spinorial 
representation. In doing this we introduce rotation 
matrices between neighboring frames which are crucial 
for the construction. In particular, using the spinorial 
representation of these matrices we are able to define 
in section \ref{secDW} the Dirac-Wilson operator. 
The standard definition of the partition function representing 
quantum amplitudes is recalled in section \ref{secSQ}.
In this section we also list the properties of the mathematical
expressions encountered in calculating the partition
function. In section \ref{secFL} we calculate the partition 
function using the hopping parameter expansion. The topological 
loop sign problem emerges naturally there. The issue of loop 
signs is discussed in more detail in section \ref{secGSO}
where the sign is defined as a function of
classes of loop homotopies. The relation between signs of
non-contractible fermionic loops and of domain-walls in 
Ising model and the topological aspect of the duality 
is discussed in section \ref{secTI}. In section 
\ref{sec2E} we give two analytic examples, 
calculating the critical temperature of the 
Ising model on the honeycomb lattice and the 
critical value of the hopping parameter on the dynamical
triangulation, making use of the existence of the
exact map between the Ising model and the fermionic
model. We close with a short discussion. 

\section{Preliminaries \label{secP}}

The aim of this paper is to discretize a theory of fermions on
a random, possibly fluctuating geometry. 
Let us first recall some basic facts about the
continuum formulation of this problem.

Consider a $D$-dimensional compact Riemannian manifold, on which
a coordinate system $\xi^\mu$ is defined. If a nonsingular
change of coordinates $\xi^\mu \to \xi^{'\mu}$ is performed at some point 
$x$ on the manifold, then a linear transformation of the com\-po\-nents
of any vector or tensor field in the tangent space 
at $x$ has also to be carried out, in order to ensure the invariance 
of the theory under coordinate transformations. For vectors, the 
matrix of this linear transformation reads~:
\begin{equation}
A^\mu_\nu(x) = \frac{\partial \xi^{'\mu}}{\partial \xi^\nu}(x) \, .
\label{Amunu}
\end{equation}
Since the change of coordinates is not singular, the determinant
of $A$ is nonzero. The matrices $A$ thus form a linear group of
non-singular real matrices $GL(D,R)$. 
The basic difficulty in any attempt
to apply the transformation law (\ref{Amunu})
to a fermionic field is that the group $GL(D,R)$ has no spinorial
representation. In other words, one cannot directly apply the information
encoded in $A$ to transform a spinor when changing the coordinates.
In order to overcome this difficulty one has to restrict somehow 
the group $GL(D,R)$ to its $SO(D)$ subgroup, which does have spinorial
half-integer representations. One can do this by introducing an additional
field of local orthonormal frames. More precisely, at each point $x$ of
the manifold one introduces a basis $e_a (x)$, $a = 1, \dots, D$, in the
tangent space, which obeys $e_a (x) \cdot e_b (x) = \delta_{ab}$
(orthonormality) and $e_1 (x) \wedge e_2 (x) \dots \wedge e_D (x) > 0$
(orientability), where the symbols $\cdot$ and $\wedge$ denote 
the internal and external products.

Expressed in a given coordinate system $\xi^\mu$, the orthonormality
and orientability conditions read~:
\begin{equation}
g_{\mu\nu} (x) \, e_a^\mu (x) \, e_b^\nu (x) = \delta_{ab} \, , \quad
e (x) \equiv \det e_a^\mu (x) = \sqrt{g(x)} > 0 \, .
\end{equation}
The matrix $e_a^\mu (x)$ is called the {\em vielbein}. It is non-singular,
and one can denote its inverse matrix by $e^a_\mu(x)$. Thus one has,
for instance, $e^a_\mu(x) e^b_\nu(x) \delta_{ab} = g_{\mu\nu}(x)$.

With these vectors one can also associate gamma matrices $\gamma^a$,
$\{ \gamma^a, \gamma^b \} = 2 \delta^{ab}$, that can be chosen so
as to have the same numerical values $\gamma^a$ for all points $x$.
One can write the Dirac matrices in the curved coordinates as
$\gamma^\mu (x) = e^\mu_a (x) \gamma^a$.

The price to pay for introducing this new field is that one also has
to introduce an additional connection on top of the Levy-Civita connection.
The new connection $\omega$ (which is called the {\em spin connection})
allows one to calculate covariant derivatives of objects that have frame
indices. For instance, the covariant derivative of the vielbein itself
is given by
\begin{equation}
\nabla_\mu e_a^\nu = \partial_\mu e_a^\nu + \Gamma^\mu_{\nu\lambda} e_a^\lambda
- \omega_{\mu a}{}^{b} \, e_b^\nu
\end{equation}
The reward is that the spin connection can be lifted to the spinorial
representation, and we can calculate the covariant derivatives of spinors
as well~:
\begin{equation}
\nabla_\mu \psi = \partial_\mu \psi + 
\frac{1}{2} \omega_{\mu ab} \sigma^{ab} \psi \, ,
\end{equation}
where $\sigma^{ab} = \frac{1}{2i} [\gamma^a, \gamma^b]$ is 
the rotation generator in the spinorial representation.

The action for fermions coupled to gravity can now be written as~:
\begin{equation}
\begin{array}{rl}
S = & \frac{1}{2} \int {\rm d}^D \xi \, e \ \bar{\psi} \, \gamma^\mu 
\nabla_\mu \; \psi \,
= \frac{1}{2} \int {\rm d}^D x \, \bar{\psi} \, (\gamma^a \cdot \nabla_a) \, \psi \\
& \\
 = & \frac{1}{2} \int {\rm d}^D x \; {\rm d}^D y \
\bar{\psi}(x) \, D(x,y) \, \psi(y) \, .
\end{array}
\label{DO1}
\end{equation}
The Dirac operator on the manifold is
\begin{equation}
D (x, y) = \delta (x - y) \, \gamma^a (x) \cdot \nabla_a (x) \, ,
\label{DO2}
\end{equation}
or, less formally, just $\gamma \cdot D$. We shall discretize this operator
in the next section. Before doing so, however, let us discuss its
topological properties.

Locally, one can always define a continuously varying field of frames.
However, doing this globally for a compact manifold is usually impossible.
What can be done instead in this case is to cover the manifold with open
patches, in each of which one can separately define a continuous field of
frames, and for any region of overlapping patches $U$ and $V$ provide
transition matrices for recalculating the frames when going from one patch
to the other~:
\begin{equation}
[e_{U}]_a (x) = [R_{UV}]_a^b \, [e_V]_b (x) \, .
\end{equation}
Here, the transition function $R_{UV}$ is a $SO(D)$ rotation matrix. It
follows that the spinors in the overlapping region can be recalculated as~:
\begin{equation}
[\psi_{U}]_\alpha (x) = [{\cal R}_{UV}]_\alpha^\beta \, 
[\psi_V]_{\beta} (x) \, .
\end{equation}
where ${\cal R}_{UV}$ is an image of $R_{UV}$ in the spinorial representation.
In a region where three patches $U, V, W$ intersect, the transition matrices
must obviously fulfill the following self-consistency equations~:
\begin{equation}
R_{UV} R_{VW} R_{WU} = \mathbbm{1} \, , \qquad
{\cal R}_{UV} {\cal R}_{VW} {\cal R}_{WU} = 
\mathbbm{1} \, .
\end{equation}
The second equation can be almost automatically deduced from the first one by
rewriting it in the spinorial representation. However, because the spinorial
representation $R \to \pm {\cal R}$ is two-valued, the signs of the
${\cal R}$'s are not automatically fixed by $R$'s. In other words, one has to
adjust in addition the signs of the transition functions for the
spinors in such a way that the consistency equation is fulfilled in any
triple intersecting patch.

This is a global topological problem. If it is solvable on the entire
manifold, the manifold is said to admit a spin structure. In two and tree
dimensions, the question of the existence of a spin structure reduces
simply to the manifold orientability; in higher dimensions the
problem is more complex.

Another important question is: how many non-equivalent spin structures are
admitted on a given manifold ? In two dimensions, the answer is $2^{2g}$,
where $g$ is the genus of the manifold \cite{gsw}. 
This number is related to the number of possible sign 
choices for independent non-contractible loops on the
manifold. 

A good discretization scheme should reflect all these topological properties.
As will be seen, the explicit construction for two-dimensional compact
manifolds to be proposed in the present paper does fulfill this requirement.

The Dirac operator (\ref{DO2}) can be expressed in local
coordinates as $\gamma^\mu \nabla_\mu$, or alternatively in frame components as
$\gamma^a \nabla_a$, $i.\,e.$ without reference to local coordinates. The
construction proposed in this paper is, in fact, coordinate-free~: we shall
express everything in frame indices $a$, without referring to coordinate
indices $\mu$.

In the lattice construction, the nearest neighbor relation that mimics the
structure of the continuum formulation will be given by a local vector~:
at each point $i$ on the dual lattice we shall define local vectors $n_{ji}$
pointing to the three neighboring vertices $j$. To calculate derivatives
(differences) in the direction of $n_{ji}$ we shall decompose it in the local
frame $e_{ia}$. Similarly, all vector,
tensor and spinor indices of objects from the tangent spaces will be expressed
in these local orthonormal frames. Lifting the construction from the 
vector to the spinor representation of the rotation group, we
shall store the information about nearest neighbors in the form of
rotation matrices. We refer to them as to the `basic rotations', 
and denote them by the
letter $B$. The advantage of using rotations is that we can express them 
in the spinorial representation, $B \to {\cal B}$.

\section{The discretization scheme \label{secDS}}

Let us start with a discussion of fermions on a regular flat lattice, using
the Wilson formulation \cite{w}. Then, we shall see how to go over, after some
modifications, to the case of irregular lattices.

The Dirac-Wilson action for free fermions reads~:
\begin{equation}
S = -\frac{K}{2} \sum_{\vec{\imath},\mu} \left\{
\bar{\Psi}_{\vec{\imath} + \vec{\mu}} (1 + \gamma^\mu) \Psi_{\vec{\imath}} +
\bar{\Psi}_{\vec{\imath}} (1 - \gamma^\mu) \Psi_{\vec{\imath} + \vec{\mu}} 
\right\} + \frac{1}{2} \sum_{\vec{\imath}}
\bar{\Psi}_{\vec{\imath}} \Psi_{\vec{\imath}} \, .
\label{sw1}
\end{equation}
where the multi-index $\vec{\imath}$ describes the node position on the
lattice, and $\vec{\mu}$ is one of the $D$ directions of the lattice. The
gamma matrices $\gamma^\mu$ are rigidly associated with these directions~:
\begin{equation}
\{ \gamma^\mu, \gamma^\nu \} = 2 \delta^{\mu\nu} \, .
\end{equation}
In the Euclidean sector, the Dirac field is represented by independent Grassmann
variables $\bar{\Psi}^\alpha$ and $\Psi^\alpha$, $\alpha = 1, \dots, N$.
In particular, for $D = 2$, the dimension of the spinor representation is
$N = 2$. In the following, spinor indices will usually be implicit; we shall
write them explicitly only when necessary.

We shall now rewrite the action (\ref{sw1}) in a coordinate-free form which
can be extended to the case of irregular lattices.

Instead of using the multi-index $\vec{\imath}$ to describe the vertex
position, we associate with each vertex a single label, say $i$, which is
a coordinate-free concept. Obviously, the particular choice of a label does
not have any physical meaning and the theory has to be
invariant under relabelings. The physical information will be encoded in
the nearest neighbor relations. 

Using these labels, the action can be cast into the following form~:
\begin{equation}
S = - K \sum_{\langle ij\rangle} \bar{\Psi}_i H_{ij} \Psi_j +
\frac{1}{2} \sum_i \bar{\Psi}_i \Psi_i \, ,
\label{sw2}
\end{equation}
where the first sum runs over oriented links connecting nearest neighbors
on the lattice. The hopping operator $H_{ij}$ is defined as
\begin{equation}
H_{ij} = \frac{1}{2}(1 + n_{ij} \cdot \gamma) \, ,
\end{equation}
where $n_{ij}$ is a local vector pointing from $j$ to $i$, being assumed that the
two are nearest neighbors. Note that in the sum over oriented links, each
link $(ij)$ appears twice, once as $\langle ij \rangle $ and once as
$\langle ji \rangle $; since we clearly have $n_{ij} = -n_{ji}$, we see
that the action (\ref{sw2}) is indeed equivalent to (\ref{sw1}).

\begin{figure}
\begin{center}
\includegraphics[height=6cm]{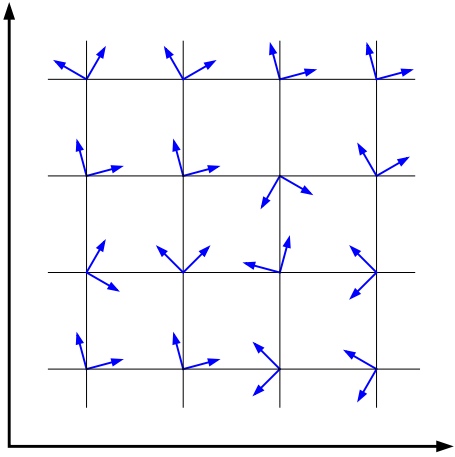}
\caption{\label{Flocframes} A hypercubic lattice with translational symmetry
and a global frame that fixes the coordinate directions for the entire
lattice. Alternatively, one can use local frames that vary from point to point.
This has the advantage of being
generalizable to a curved background.}
\end{center}
\end{figure}

Even at this stage it is more elegant to stop referring to
coordinates and instead use components of the global frame
$E_a = (E_1, E_2)$. Thus, we replace $\gamma^\mu$ by $\gamma^a$,
and decompose the nearest neighbor vector $n_{ji}$ into components in
this frame. The product $n_{ij} \cdot \gamma$ can then be expressed as~:
\begin{equation}
n_{ij} \cdot \gamma = n_{ij, a} \gamma^a
= n_{ij, 1} \gamma^1 + n_{ij, 2} \gamma^2 \, .
\end{equation}
Written in the form (\ref{sw2}), the action is now coordinate-free, but
it still depends on the global frame through the vector components
$n_{ij, a}$ and the gamma matrices $\gamma^a$. Such a global
frame and a common spinorial basis exist only in exceptional geometries, like
the regular torus or plane. In order to define a theory on another topology
or, generally, on a curved background, we have to get rid of this concept
and use local frames instead. 

One can introduce independent orthonormal frames as in fig. \ref{Flocframes}.
At each lattice point $i$ one has a pair of orthonormal vectors
$(e_{i1}, e_{i2})$. In particular, on a torus the local frames $e_{ia}$
can be obtained from the global frame $E_a$ by local rotations~:
\begin{equation}
e_{ia} = \left[R_i\right]_a^b E_b \, .
\label{eE}
\end{equation}
The spinor components $\Psi_i$ are transformed by these rotations into their
components in the local bases $\psi_i$~:
\begin{equation}
\psi_{i\alpha} = \left[{\cal R}_{i}\right]_\alpha^\beta \Psi_{i\beta} \quad , 
\quad \bar{\psi}_i^\alpha = \bar{\Psi}_i^\beta 
\left[{\cal R}_i^{-1}\right]_\beta^\alpha \, ,
\label{pP}
\end{equation}
where the matrices ${\cal R}_i$ belong to the half-integer representation of the
rotations $R_i$~:
\begin{equation}
{\cal R}_i \gamma^a {\cal R}^{-1}_i
= \left[R_i\right]^a_b \, \gamma^b \, .
\label{repr}
\end{equation}
In component-free notation the equations (\ref{eE}), (\ref{pP}) and
(\ref{repr}) read~:
\begin{equation}
e_i = R_i E \, , \quad
\psi_i = {\cal R}_i \Psi_i \, , \quad
\bar{\psi}_i = \bar{\Psi}_i {\cal R}_i^{-1} \, , \quad
{\cal R}_i \gamma {\cal R}^{-1}_i = R_i \gamma \, .
\end{equation}
Using this notation, one should remember that the matrix ${\cal R}$
acts on the spinor indices whereas $R$ acts on the frame indices. Using
the local frames, we can write the action (\ref{sw2}) as~:
\begin{equation}
S = - K \sum_{\langle ij \rangle} \bar{\psi}_i {\cal H}_{ij} \psi_j +
\frac{1}{2} \sum_i \bar{\psi}_i \psi_i
\label{sh}
\end{equation}
where 
\begin{equation}
{\cal H}_{ij} = {\cal R}_i H_{ij} {\cal R}^{-1}_j 
= \frac{1}{2} {\cal R}_i \left[ 1 + n_{ij} \cdot \gamma \right]
{\cal R}_i^{-1} \underbrace{{\cal R}_i {\cal R}_j^{-1}}_{{\cal U}_{ij}} \, .
\label{hu}
\end{equation}
Here, ${\cal U}_{ij}$ is a matrix allowing to recalculate the
components of a spinor going from a frame $j$ to the frame $i$.
In other words, it is a sort of a connection matrix that performs a parallel
transport of spinors between neighboring vertices.

So far, equation (\ref{hu}) is written in a hybrid notation, because
the spinors are already expressed in the local frames $e_i$ whereas $n_{ij}$
and $\gamma$ are still written in the global frame $E$. However, 
applying (\ref{repr}) to (\ref{hu}) one finds~:
\begin{equation}
{\cal R}_i \; n_{ij} \cdot \gamma \; {\cal R}^{-1}_i =
n_{ij,a} \, {\cal R}_i \gamma^a {\cal R}_i^{-1} = n_{ij,a} R^a_b \, 
\gamma^b = n_{ij}^{(i)} \cdot \gamma
\label{glob_loc}
\end{equation}
where in the local basis the vector $n_{ij}^{(i)}$ has the components
\begin{equation}
n_{ij,b}^{(i)} = n_{ij,a} R^a_b  \, ,
\end{equation}
different from the global frame components $n_{ij,a}$ .
The new bracketed index $(i)$ now differentiates between different local
frames where the components of the vector are calculated; thus,
$n^{(i)}_{ij}$ refers to the same vector as $n^{(j)}_{ij}$, but with 
components expressed in a different frame. Intuitively, what the equation
means is simply that the components of a vector in a rotated basis can be
alternatively calculated by performing the inverse rotation on the vector
itself while keeping the basis fixed.

An important point is that the crossover from the global description to the
local one as in (\ref{glob_loc}) preserves the numerical values of the
$\gamma^a$ matrices. In other words, $\gamma^1$ associated with the local
direction $e_{i1}$ at a point $i$ has the same numerical value as
$\gamma^1$ associated with the $e_{j1}$ at any other point $j$, and
likewise for $\gamma^2$.

Using the components $n_{ij}^{(i)}$ of the nearest neighbor vector in the
local frame $i$, we can now write (\ref{hu}) as
\begin{equation}
{\cal H}_{ij} = \frac{1}{2} \left[ 1 + n^{(i)}_{ij} \cdot \gamma \right]
\ {\cal U}_{ij} \, .
\label{huprim}
\end{equation}
Alternatively, using the features of $n_{ij}^{(i)}$ discussed
above, we can cast the hopping operator into several equivalent forms~:
\begin{equation}
{\cal H}_{ij} = \frac{1}{2} 
\left[ 1 + n^{(i)}_{ij} \cdot \gamma \right] {\cal U}_{ij} =
\frac{1}{2} \left[ 1 - n^{(i)}_{ji} \cdot
\gamma \right] {\cal U}_{ij} =
{\cal U}_{ij} \frac{1}{2} \left[ 1 + n^{(j)}_{ij} \cdot \gamma \right] \, .
\label{hopping}
\end{equation}
These different expressions for ${\cal H}_{ij}$ correspond to different
ways of calculating the hopping term
$\bar{\psi}_i {\cal H}_{ij} \psi_j$ in (\ref{sh}). One method is to first
parallel transport the spinor $\psi_j$ from $j$ to $i$, getting
${\cal U}_{ij} \psi_j$, and then to calculate the corresponding scalar
in the frame $i$, as is done on the left hand side of (\ref{hopping}).
Sometimes it is convenient to replace $n_{ij} = -n_{ji}$ in order to
change the direction of the vector between indices $i$ and $j$, as is done
in the second expression. Alternatively, one can first transport the spinor 
$\bar{\psi}_i$ from $i$ to $j$~, which gives $\bar{\psi}_i {\cal U}_{ij}$,
and then calculate the corresponding scalar in the frame $j$, as is done
on the right hand side, {\em etc}. All these expressions are equivalent
and can be deduced from each other, so that the most convenient one is
always chosen.

The additional upper index in the brackets makes formulae visually less
transparent but removes the logical ambiguity which otherwise might lead
to confusion. We will therefore extend this notation to all objects
occurring in our construction. For example,
$\psi_j^{(i)} = {\cal U}_{ij} \psi_j^{(j)}$ means that the spinor
$\psi_j$ is transported from $j$ to $i$. Similarly,
$\bar{\psi}_i^{(j)} = \bar{\psi}_i^{(i)} {\cal U}_{ij}$
means that $\bar{\psi}_i$ is transported from $i$ to $j$.
There is no summation over the repeated indices. 
The only exception will be made for objects calculated in the
frame belonging to the point where they are themselves defined, 
since in this case
leaving out the upper index does not cause any ambiguity. For example, we
will write $\psi_i$ instead of $\psi_i^{(i)}$.

Using this notation, the Wilson action becomes~:
\begin{eqnarray}
S & = & - K \sum_{\langle ij \rangle} \bar{\psi}_i
\frac{1}{2} \left[ 1 + n^{(i)}_{ij} \cdot \gamma \right] 
\psi^{(i)}_j +
\frac{1}{2} \sum_i \bar{\psi}_i \psi_i \ . 
\label{definition}
\end{eqnarray}
Contrary to (\ref{sw1}), this form of the Wilson action can now be
generalized to any random irregular lattice. It also makes direct contact
with the continuum formalism (\ref{DO1}). Finally, note that it is invariant
under a change of the local frames~:
\begin{equation}
e_i \to R_i e_i \, , \quad 
\Psi_i \to {\cal R}_i \Psi \, , \quad
\bar{\Psi}_i \to \bar{\Psi}_i {\cal R}_i^{-1} \, , \quad
{\cal U}_{ij} \to {\cal R}_i {\cal U}_{ij} {\cal R}_j^{-1} \, .
\label{inv}
\end{equation}
where $R_i$ are arbitrary local rotations, and ${\cal R}_i$
are the corresponding matrices in the spinorial representation.

\section{A topological problem \label{secTP}}

Let us return to the consequences of the fact that the (spinorial)
half-integer representation of the rotation group is actually only a
representation up to a sign factor.

In two dimensions, the $SO(2)$ group can be parametrized by a
single parameter $\phi \in [0,2\pi)$. For a given value of this parameter 
the rotation matrix is given by~:
\begin{equation}
R(\phi) = e^{\phi \epsilon} = \cos(\phi) + \epsilon \sin(\phi)
= \left( \begin{array}{rr} \cos(\phi) & \sin(\phi) \\
                          -\sin(\phi) & \cos(\phi)
         \end{array}\right)
\label{Rphi}
\end{equation}
where $\epsilon_a^{~b}$ is the standard antisymmetric matrix with
$\epsilon_1^{~2} = 1$.

The corresponding matrix ${\cal R}(\phi)$ in the spinorial representation is
${\cal R}(\phi) = e^{\frac{i}{2} \sigma^{12} \phi}$. In particular, if
we set $\gamma^1 = \sigma_3$ and $\gamma^2 = \sigma_1$, where $\sigma_i$
are the Pauli matrices, then $\sigma^{12} = \sigma_2$ and 
rotation matrix is~:
\begin{equation}
{\cal R}(\phi) = e^{\frac{i}{2} \sigma_{2} \phi}
= \cos(\phi/2) + \epsilon \sin(\phi/2) 
= \left( \begin{array}{rr} \cos(\phi/2) & \sin(\phi/2) \\
                          -\sin(\phi/2) & \cos(\phi/2)
         \end{array}\right)
\label{Ophi}
\end{equation}
where $\epsilon = i \sigma_2$ is an antisymmetric tensor that is
numerically identical with the one in (\ref{Rphi}). The difference, of
course, is that the tensor in equation (\ref{Rphi}) has frame indices
$\epsilon_a^{~b}$ whereas the one in (\ref{Ophi}) has spinorial indices 
$\epsilon_\alpha^{~\beta}$.

In order to fix the global sign of ${\cal R}(\phi)$, on should
control the angle $\phi$ in the range $[0, 4\pi)$ rather than the usual
$[0, 2\pi)$. This would require changing continuously the
angle and calculating the overall change $\int \! d\phi$ keeping track of the
number of `full circles'. However, this cannot be done here since 
the relative angles between the frames $e_{ia}$ are 
determined in the fundamental range $[0, 2\pi)$ only.

\begin{figure}
\begin{center}
\includegraphics[width=6cm]{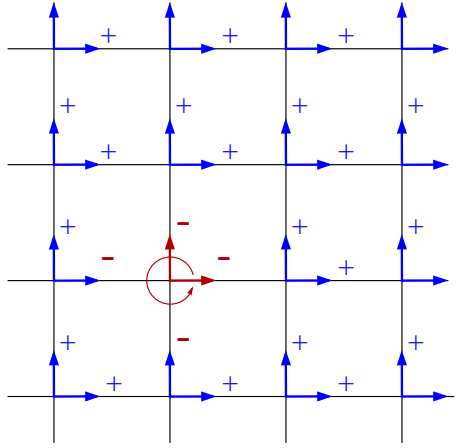}
\caption{\label{Fsign_topo}Rotation of a local frame by $2\pi$. Even
though the resulting frame configuration is obviously the same as before,
spinor components can change their sign due to the sign ambiguity.}
\end{center}
\end{figure}

The sign ambiguity also has topological consequences. Consider once more
the regular, toroidal, flat lattice and choose on it a constant field of
identical frames (see fig. \ref{Fsign_topo}). We first set
${\cal U}_{ij} = \mathbbm{1}$ for all links. Trivially, if
at a vertex $i$ the frame is rotated by $2 \pi$, the frame
configuration does not change. However, because  
${\cal R}_i (2 \pi) = -\mathbbm{1}$
in the spinorial representation, all links emerging from $i$ acquire
a negative sign ${\cal U}_{ji} = -\mathbbm{1}$ according to the
transformation law (\ref{inv}). The resulting `sign field' is different
from the original one but at the same time equivalent to it. By repeating
this procedure in other vertices one can produce many different, but equivalent, 
sign configurations for the same field of frames.

It is easy to see that a local rotation of a frame by $2 \pi$ preserves
the overall sign of all elementary plaquettes, {$i.\,e.$} the product
of signs of all links on the plaquette's perimeter. Thus, for any
configuration obtained from the original one, all elementary plaquettes
have a positive overall sign. We shall require this to be true in general,
$i.\,e.$ for any configuration of local frames on the lattice the sign of
all elementary plaquettes is set to $+1$; this ensures that spinors
remain unchanged by parallel transport around any elementary plaquette.
This requirement is dictated by the underlying continuum theory, in which
parallel transport of a spinor around a closed loop in a locally flat
patch leaves the spinor intact. Later on, for curved lattices, we shall
modify this constraint so as to adjust it to the case where there is a
deficit angle inside an elementary plaquette.

Assuming that all elementary plaquettes have a positive sign
we can prove now some simple topological theorems concerning the signs of
loops on the lattice.

\begin{figure}
\begin{center}
\psfrag{ll}{{\footnotesize $L$}}
\psfrag{lp}{{\scriptsize $L\cap P$}}
\psfrag{pp}{{\footnotesize $P$}}
\psfrag{l+p}{{\footnotesize $L\cup P$}}
\includegraphics[height=5.5cm]{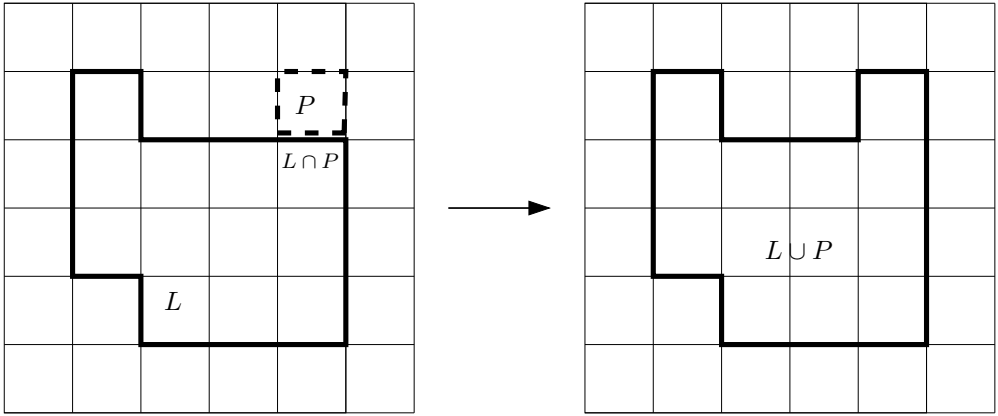}
\caption{\label{Fdefo1} A small deformation of a loop $L$ (bold line)
by an elementary loop $P$ (dashed line), resulting in the loop $L'$.}
\end{center}
\end{figure}

It is convenient to define an auxiliary operation for loops on a lattice,
to be called a {\em small deformation} of a loop. To deform a
loop $L$, we pick an elementary plaquette $P$ which shares at least one common
link with $L$, and substitute the intersection $L \cap P$ by the
complementary part of $P$, resulting in a new loop $L' = L\cup P - L\cap P$
(see fig. \ref{Fdefo1}).\footnote{Somewhat more precisely, we also have to
require that the intersection $L \cap P$ be connected, so as to avoid 
situations in which a small deformation splits a loop into two or
more parts.}

As with elementary plaquettes, we can define the overall sign of a loop
as the product of signs of all links on the loop. One easily checks that
the sign of the deformed loop $L'$ is the same as that of $L$ -- namely,
the addition of $P$ to $L$ cannot change the sign because $P$ has a
positive sign by default, and the removal of the intersection $L \cap P$
cannot change the sign because each link is `removed twice' (once from
$P$ and once from $L$), so that the total number of removed links is always
even.

Any contractible loop can be obtained from the elementary loop by a sequence 
of small deformations. Thus all contractible loops have positive signs.

\begin{figure}
\begin{center}
\includegraphics[height=6cm]{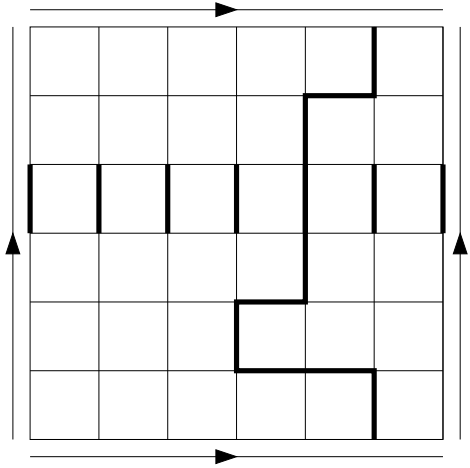}
\caption{\label{Fyloop} A non-contractible loop on a toroidal lattice
with a constant frame. The single links drawn as bold lines all have
transition matrices ${\cal U}_{ji} = -\mathbbm{1}$, whereas all other
links have ${\cal U}_{ji} = \mathbbm{1}$; as a consequence, the loop
has a negative overall sign.}
\end{center}
\end{figure}

This is not, however, the case with non-contractible loops, which can
take either sign. An example of a loop with negative sign is shown in
fig. \ref{Fyloop}~: if we choose ${\cal U}_{ji} = -\mathbbm{1}$ for one
complete row of links on the lattice (as in the figure)
and ${\cal U}_{ji} = \mathbbm{1}$ everywhere
else, then any loop that encircles the lattice in the $y$ direction passes
through exactly one link with negative sign, and thus has a negative overall sign
 \footnote{More generally, if a loop which encircles
the lattice in the $y$ direction goes back and forth having a sort of
$S$ shape, it may cross links with negative signs more than once. The
number of crossings is however odd.}. 

Obviously, two sign configurations are equivalent if one can transform one
into the other by a sequence of local rotations
${\cal R}_i (2 \pi) = -\mathbbm{1}$. Because local rotations do not change
the sign of any loop, a configuration with at least one loop of negative
sign cannot be equivalent to a configuration that has only loops of positive
sign. In other words, the two sign configurations are topologically distinct. 

Now, using small deformations we can easily prove that all non-contractible
loops encircling the torus in the same direction must have the same sign. 
This means, for example, that it is sufficient to calculate the sign of just
one `vertical' loop (which encircles the lattice in the $y$ direction) to
know the sign of all other vertical loops. More generally, the sign of a
loop is not a property of a single loop but rather of all loops in the same
homotopy class, $i.\,e.$ those that can be obtained from each other by a
sequence of small deformations. On the torus there are two independent
non-trivial homotopy classes of loops (`vertical' and `horizontal') and,
therefore, four distinct possible sign configurations. These, in turn, correspond
to four distinct spin structures.

The statement can be generalized by observing that there are $2g$
independent classes of non-contractible loops on a surface with genus $g$,
which means that there are $2^{2g}$ different sign configurations and thus the
same number of spin structures. In particular, a lattice with spherical
topology admits only one spin structure.

\begin{figure}
\begin{center}
\includegraphics[height=5.5cm]{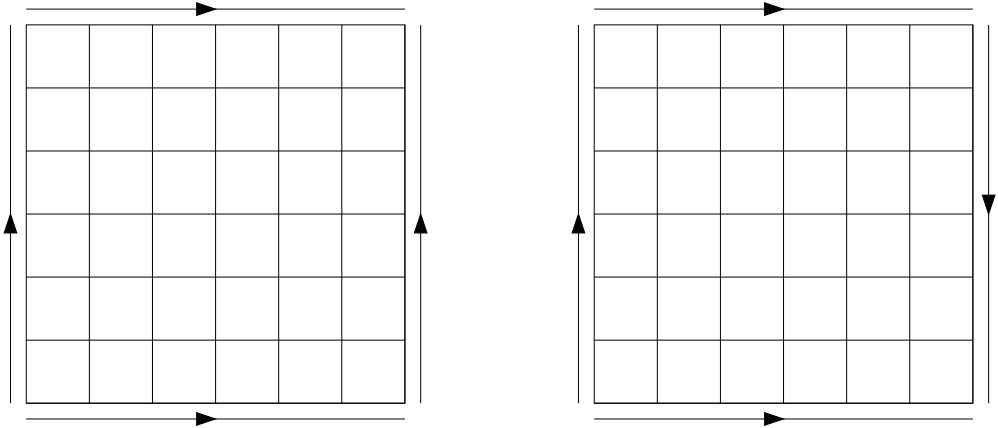}
\caption{\label{FKbottle} (Left) A lattice with toroidal boundary conditions.
(Right) A lattice with the boundary conditions of a Klein bottle. The arrows
indicate the directions in which the opposite edges are to be taken when
joined together.}
\end{center}
\end{figure}

On the other hand, on a non-orientable lattice one cannot globally define
a field of orientable frames. An example of such a lattice is the so-called
one-sided torus or Klein bottle, which is constructed in the same way as the
standard torus but has different boundary conditions, as shown in
fig. \ref{FKbottle}. It is possible to show that a frame transported along a
closed path would have changed its
handedness after a complete tour around the lattice. 
Because there does not exists a field of orientable frames, one 
cannot in this case define a spin structure or a Dirac operator.

\section{Local frames on a random lattice \label{secLF}}

The form (\ref{sw1}) of the Wilson action is particularly simple not only
because of the simple topology of the torus, which allows for the definition
of a global frame, but also because of the regular geometry of the lattice
which everywhere repeats the same simple motif. On an irregular lattice,
local angles and link lengths change from point to point. This must be
reflected in the construction of the hopping term, which depends on these
local details through the covariant derivative.

To make the geometrical part of the discussion as simple as possible, and
to minimize the number of local degrees of freedom of the lattice, we 
restrict the discussion to equilateral random triangulations. This 
greatly reduces the number of local degrees of freedom, making the
discussion more transparent and allowing us to focus on the interesting
topological part of the problem. Let us mention, however, that the presented
construction can be easily generalized to the case of variable
link lengths and angles.

On an equilateral triangulation, the local geometry is completely encoded
in the connectivity of the lattice; all other details are fixed by the
simple geometry of the equilateral triangle. In particular, the deficit
angle at a vertex $i$ is determined solely by its order $q_i$ : 
$\Delta_i = (6 - q_i) \pi/6$.

The local curvature of the lattice is concentrated in the vertices of the
triangulation. The geometry becomes singular in these points and therefore
it is difficult to provide a unique definition of a tangent space at the
vertices. It is more convenient to define tangent spaces at the dual points
of the lattice, $i.\,e.$ at the centers of the triangles. Inside each
triangle the geometry is locally flat and thus naturally spans a tangent
space. We therefore locate all local frames, and also all fermionic
fields, at the centers of the triangles. Each point $i$ where a field is
defined has then three neighbors, each of which at the same
distance from $i$. The vectors pointing to the neighbors are also 
equally spaced in the angular variable, $i.\,e.$ they are
separated by angles $2 \pi / 3$.

\begin{figure}
\begin{center}
\includegraphics[height=7cm]{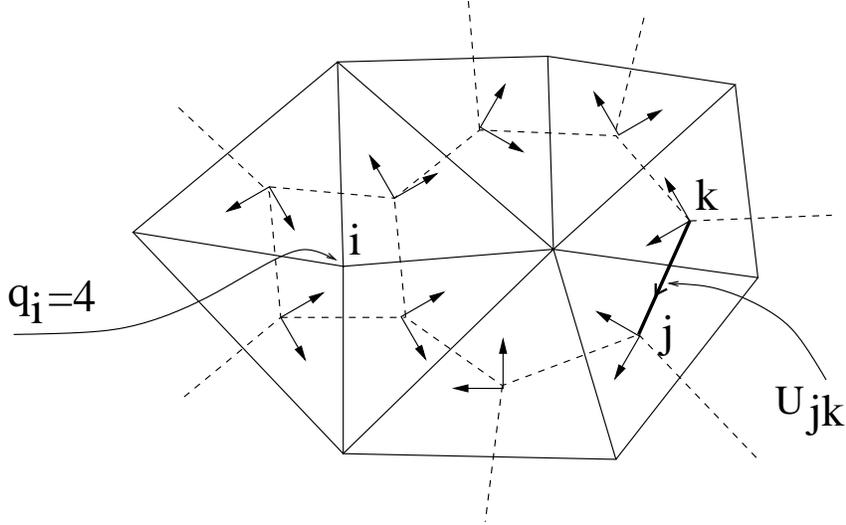}
\caption{\label{FDTframes} A small piece of a random triangulation with
local frames. $U_{jk}$ is the transition matrix between the frames at $k$
and $j$, and $q_i$ is the order of the vertex $i$.}
\end{center}
\end{figure}

Before defining the fermionic fields, however, let us discuss the properties
of the field of oriented orthonormal local frames on such a random
triangulation. An example of a triangulation decorated with frames is shown
in fig. \ref{FDTframes}.

At each triangle $i$ live two orthonormal vectors $e_{i1}$ and $e_{i2}$
such that $e_{ia} \cdot e_{ib} = \delta_{ab}$. Apart from the internal
product there is also an external one $\wedge$, which enables one to choose
frames with the same handedness $e_{i1} \wedge e_{i2} > 0$ for all triangles.
Now consider two neighboring triangles $i$ and $j$, each endowed with its
own frame $e_i$ and $e_j$. The interiors of the two triangles together form
a flat patch of the triangulation. One can think of the two frames as being
two alternative frames for the same patch. One can calculate components of
our objects in either one of them, and easily recalculate them when going from
one to the other. To this purpose introduce $SO(2)$ transition
matrices $U_{ij}$ and $U_{ji}$ such that~:
\begin{equation}
U_{ij} U_{ji} = \mathbbm{1} \, , \quad 
e_i = U_{ij} e_j \, , \quad
e_j = U_{ji} e_i \, .
\label{uu}
\end{equation}
One can repeat the same calculation for any pair of neighboring triangles
and use it to transport a frame between any two points $i_1$ and $i_n$
along an open path $C = (i_1, i_2, \dots, i_n)$~:
\begin{equation}
e_{i_n} = U_{i_{n}i_{n\!-\!1}}\, 
\dots \, U_{i_3i_2} \, U_{i_2i_1} \, e_{i_1} = U(C) \ e_{i_1} \, .
\end{equation}
Since we study a theory whose content is independent of the choice of
frames, we are interested in the pertinent transformation laws  and 
in quantities invariant under local
$SO(2)$ rotations of the frames~: $e_i \to e'_i = R_i e_i$.
The object $U (C_{ji}) = U_{jk} \dots U_{ni}$ for any open path between $i$
and $j$ transforms as~:
\begin{equation}
U (C_{ji}) \to U' (C_{ji}) = R_j U (C_{ji}) R^{-1}_i \, ,
\end{equation}
as one can see from (\ref{uu}). In particular, for a closed path $L_i$
beginning and ending at the same triangle $i$, $U (L_i)$ transforms as
\begin{equation}
U (L_i) \to U' (L_i) = R_i U (L_i) R_i^{-1} \, ,
\end{equation}
and hence ${\rm Tr} \, U (L_i)$ is an invariant. Moreover, this invariant
does not depend on the choice of the initial point $i$ of the loop, and
is thus a property of the loop $L$ itself. It is a geometrical quantity
related simply to the total angle $\int {\rm d} \alpha$ by which
a tangent vector is rotated when transported along the loop. On a flat
lattice, this angle is a multiple of $2 \pi$. On a curved lattice the
situation is somewhat more complicated. In particular, for an
elementary loop $L_q$ surrounding a vertex of order $q$, the loop
invariant is
\begin{equation}
\frac{1}{2} {\rm Tr} \, U (L_q) = 
\cos \frac{q \pi}{3} = \cos \frac{(6 - q) \pi}{3}
= \cos \Delta_q
\label{loop}
\end{equation}
and contains information about the deficit angle $\Delta_q$, or equivalently
about the curvature at the vertex. There are various possibilities to prove
this statement; the proof outlined here offers us an opportunity to introduce
an auxiliary construction which will be useful throughout the remaining
part of the paper, especially when we shall lift the spin connection to the
spinorial representation.

Recall that the information about the local geometry of the lattice is
stored in the form of three local unit vectors $n^{(i)}_{ji}$ pointing
from $i$ to its three nearest neighbors. There is, however, another and
for the problem at hand more suitable way of achieving 
the same goal. Instead of the vectors
$n^{(i)}_{ji}$ themselves, one can equivalently consider the rotations that
connect $n^{(i)}_{ji}$ to $e_i$. To introduce the rotation matrices,
we first associate an entire frame with each of the three nearest neighbor
vectors, treating $n^{(i)}_{ji}$ as the first basis vector of each
corresponding frame. The second base vector of the frame is then
automatically determined by the orthonormality condition. Now we have three
particular frames $n^{(i)}_{ji,a} = (n^{(i)}_{ji,1}, n^{(i)}_{ji,2})$ for
the three neighbors $j$ of $i$. The frames $n^{(i)}_{ji}$ can be obtained
from the local frame $e_i$ by a rotation $B^{(i)}_j$~:
\begin{equation}
n^{(i)}_{ji} = B^{(i)}_j e_i \ .
\end{equation}
We refer to them as to the {\em basic rotations} at $i$.

\begin{figure}
\begin{center}
\psfrag{i}{$i$}
\psfrag{j}{$j$}
\psfrag{ei}{$e_{i1}$}
\psfrag{ej}{$e_{j1}$}
\psfrag{n}{$n_{ji}$}
\psfrag{m}{$n$}
\psfrag{k}{$k$}
\psfrag{B}{$$}
\includegraphics[height=6cm]{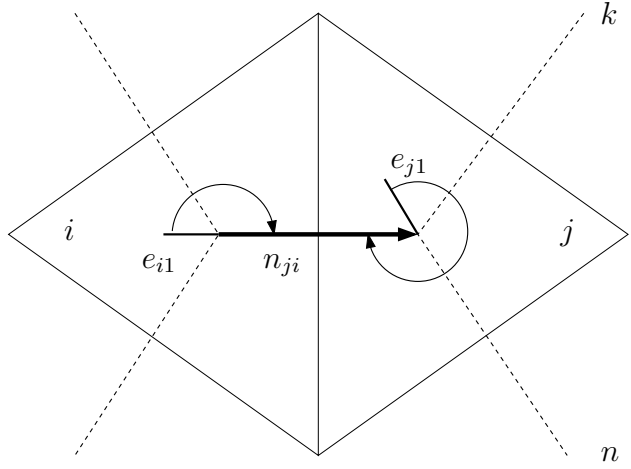}
\caption{\label{FBrot} A patch of two neighboring triangles, and the three
nearest neighbor vectors $n_{ij}$ for each of them. The same information
can be provided by a rotation matrix between $n_{ij}$ and the first basis
vector $e_{j1}$, shown as the flag emerging from the center of each triangle.
In this example, the basic rotation $B^{(j)}_i$ of frame $j$ to the direction
of its neighbor $i$ is a rotation by $5 \pi / 3$, whereas the basic rotation
$B^{(i)}_j$ of frame $i$ to the direction of its neighbor $j$ is a rotation
by $\pi$.}
\end{center}
\end{figure}

Now, it is convenient to decompose the connection matrices $U_{ji}$ into
basic rotations $B^{(i)}_j$ at $i$ and $B^{(j)}_i$ at $j$. Letting them act
first on the frame $e_i$, one obtains the frame $n^{(i)}_{ji}$. One then
flip it to the frame $n^{(j)}_{ij}$ using a rotation by $\pi$, which is
represented by the matrix $F = e^{\epsilon \pi}$. Finally, using the inverse
basic rotation at $j$ one rotates it to $e_j$. In other words, the transition
from $e_i$ to $e_j$ (and vice versa) can be done in the following three steps
(see fig. \ref{FBrot})~:
\begin{equation}
e_j = [B^{(j)}_i]^{-1} F B^{(i)}_j \, e_i \, , \quad
e_i = [B^{(i)}_j]^{-1} F B^{(j)}_i \, e_j \, .
\end{equation}
Comparison with (\ref{uu}) leads to~:
\begin{equation}
U_{ij} = [B^{(i)}_j]^{-1} F B^{(j)}_i \, , \quad
U_{ji} = [B^{(j)}_i]^{-1} F B^{(i)}_j \, .
\label{bfb}
\end{equation}
One can use this decomposition to calculate the loop invariants
${\rm Tr} \, U (L)$~:
\begin{eqnarray}
{\rm Tr} \, U(L) & = &
{\rm Tr} \, \prod_{k=1}^n U_{i_{k+1}i_k} =
{\rm Tr} \, \prod_k T_{i_k}  
\label{ttt}
\end{eqnarray}
where $\prod$ is an ordered product that runs through all vertices
on the loop $L = (i_1, i_2, \dots, i_n)$ with the cyclic boundary condition
$i_{n+1} = i_1$ and the rotation matrices 
\begin{equation}
T_{i_k} \equiv B^{(i_k)}_{i_{k+1}} [B^{(i_k)}_{i_{k-1}}]^{-1} F
= e^{(\pm)_{i_k} \frac{\pi}{3} \epsilon}
\label{turn}
\end{equation}
correspond to the turn taken by the path at the triangle $i_k$ \cite{kw}. 
It depends on the turn-angle, which 
can be either $+\pi/3$ if the path turns to the
left or $-\pi/3$ if it turns to the right. In fact, on
a equilateral triangulation, the sign $(\pm)_{i_k}$ 
determines completely the turn matrix $T_{i_k}$
at the triangle $i_k$. It does not depend on the
particular orientation of the frame, because under 
rotation of the frame $i_k$
the basic rotations transform as~:
\begin{equation}
B^{(i_k)} \to B^{(i_k)} R_{i_k} \, , \quad
[B^{(i_k)}]^{-1} \to R_{i_k}^{-1} [B^{(i_k)}]^{-1}
\end{equation}
thus leaving the combination $B^{(i_k)} [B^{(i_k)}]^{-1}$ in $T_{i_k}$ intact.

An elementary loop around a vertex of order $q$ turns exactly $q$ times in
the same direction. Thus we have
\begin{equation}
\frac{1}{2} {\rm Tr} \, U (L_q) = {\rm Tr} \ e^{\pm \frac{q \pi}{3}\epsilon} =
\cos \frac{q\pi}{3} = \cos \frac{(6-q)\pi}{3} \, .
\label{lq}
\end{equation}
as claimed in (\ref{loop}).

\section{The spinorial representation \label{secSR}}

The next step is to lift the connections $U_{ij}$ to the spinorial
representation, $U_{ij} \to {\cal U}_{ij}$. We continue to use the
convention of denoting all rotation matrices in the spinorial representation
by calligraphic letters~: $U \to {\cal U}$ for connections, $B \to {\cal B}$
for basic frame rotations, $T \to {\cal T}$ for turns and $F \to {\cal F}$
for flips.

\begin{figure}
\begin{center}
\includegraphics[height=6cm]{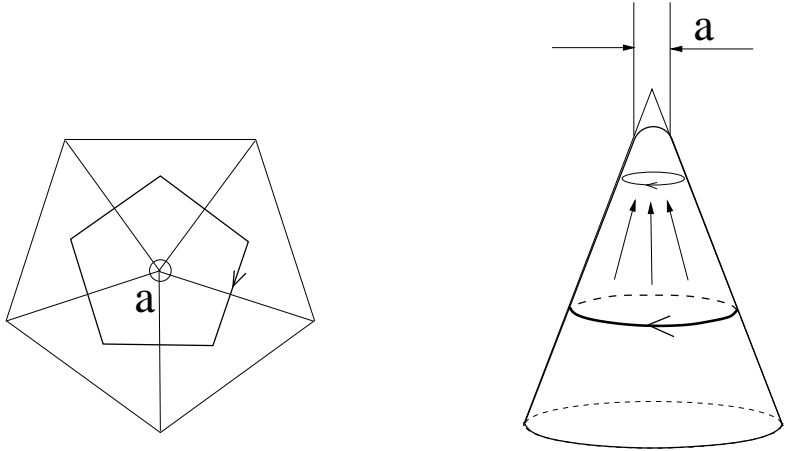}
\caption{\label{FRcone} The internal geometry of a set of triangles around
a vertex is the same as that around the peak of a cone~: it is flat
everywhere except for a single point where the curvature is concentrated
in a singularity. We can determine the sign of any loop around the cone if
we first regularize this singularity by `flattening' the cone, and find
$S = +1$.}
\end{center}
\end{figure}

The starting point of the construction is the decomposition (\ref{bfb}).
If we write it in the spinorial representation, each matrix that occurs in
this equation is determined only up to a sign~: 
$e^{\epsilon \phi} \to \pm e^{\epsilon \phi/2}$ (\ref{Ophi}). The idea
is now to affix the spinorial representation of all matrices on the right
hand side of (\ref{bfb}) with a positive sign~:
\begin{eqnarray}
B= e^{\epsilon \phi} & \to & {\cal B} = e^{\epsilon \phi/2} \\
F= e^{\epsilon \pi} = \mathbbm{1} & \to & {\cal F} = e^{\epsilon \pi/2}
= \epsilon \, ,
\label{srep}
\end{eqnarray}
and keep the sign $s_{ji} = \pm 1$ as a separate variable for each link~:
\begin{equation}
U_{ij} \to {\cal U}_{ij} = 
s_{ij} \, [{\cal B}^{(i)}_j]^{-1} \epsilon \, {\cal B}^{(j)}_i \, , \quad
U_{ji} \to {\cal U}_{ji} = 
s_{ji} \, [{\cal B}^{(j)}_i]^{-1} \epsilon \, {\cal B}^{(i)}_j \, .
\label{ss1}
\end{equation}
We demand that parallel transport of a spinor along a given link and back
does not change the spinor. We see that this is indeed the case, $i.\,e.$
we have ${\cal U}_{ji} {\cal U}_{ij} = \mathbbm{1}$ if
\begin{equation}
s_{ji} s_{ij} = -1 \, .
\label{ss}
\end{equation}
Using a similar calculation as the one which led
to (\ref{lq}) one finds that in the spinorial
representation the loop invariant for 
an elementary loop around a vertex is
\begin{equation}
\frac{1}{2} {\rm Tr} \, {\cal U} (L_q) = 
S_{L_q} \cdot \cos \frac{\Delta_q}{2} \, .
\end{equation}
where $\Delta_q$ is the deficit angle, and $S_{L_q}$ is 
a sign $\pm$. The factor one-half in the argument 
of the cosine follows from (\ref{srep}).
The total sign of the loop, denoted by $S_{L_q}$,
depends on the choice of signs $s_{ij}$ in (\ref{ss1}) and
has to be calculated. We require that the signs $s_{ij}$
are chosen in such a way that for each elementary
loop the sign $S_{L_q}$ is positive~:
\begin{equation}
S_{L_q} = 1 \, .
\label{ss2}
\end{equation}
Note that for $q = 6$ this requirement is natural,
because the plaquette is flat,
$\Delta_6 = 0$, and as discussed before for a flat patch
the parallel transport should be trivial~:
${\cal U} (L_6) = \mathbbm{1}$. Thus indeed we should
have $S_{L_6}=1$. Also for other $q$'s the requirement can
be motivated. The geometry of an elementary plaquette corresponds to
the geometry of a flat cone, which has a singularity at the peak. The
elementary loop encircles this singularity at some distance $r$ from the
peak. One can regularize the singularity by smoothing the peak, $i.\,e.$
replacing it by a differentiable surface (see fig. \ref{FRcone}).
 
In doing so, one deforms only a very small region within a distance of
$\epsilon$ around the peak, where $\epsilon \ll r$. 
Now imagine that we shrink the loop, continuously decreasing 
its radius. Then ${\rm Tr} \, {\cal U} (r)$ and $\Delta (r)$ 
both change continuously with $r$. In the limit $r \to 0$, 
the loop ends up on the top of the regularized part of 
the geometry which is flat. Thus, again $S=+1$ in the limit 
of $r \to 0$. This already is sufficient to have positive sign 
for all values of $r$, because in the course of continuous 
changing, the deficit angle $\Delta$ was changing continuously
and hence the sign $S$ could not have jumped between negative to positive 
values without making ${\cal U}$ discontinuous. In other
words, $S$ must keep the value $+1$ for all $r$.

Because the regularized zone can be made arbitrarily small,
we assume that the triangulated lattice, which 
corresponds to the limit $\epsilon\rightarrow 0$,
inherits the property of the regularized geometry:
 the sign of any elementary loop is $S_{L_q} = +1$ 
for any $q$.

In order to enforce the constraint $S_{L_q}=+1$ for
each plaquette, one has to establish a relation between 
$S_{L_q}$ and the signs of links $s_{ji}$. 
In analogy to (\ref{ttt}), one can calculate the
loop invariant in the spinorial representation as~:
\begin{equation}
{\rm Tr} \, {\cal U}(L) = 
{\rm Tr} \, \prod_{k=1}^n {\cal U}_{i_{k+1}i_k} 
= \prod_k s_{i_{k+1}i_k} \, \cdot \, {\rm Tr} \, \prod_k {\cal T}_{i_k}
\, .
\label{sttt}
\end{equation}
Comparing this to the result pertinent for the
fundamental representation (\ref{ttt}), one finds that
an additional product of link signs appears, as expected. But there
is also another source of signs hidden in (\ref{sttt}). 
It has its origin in the spinorial representation of the turn 
matrices $T\rightarrow {\cal T}$. Surprisingly, and in contrast to
the fundamental representation, the product of basic
rotations depends on the position of the frame. More precisely, 
calculating the rotation corresponding to the turn taken by the path at
$i_k$ one gets an additional sign $z_{i_k}$~:
\begin{equation}
{\cal T}_{i_k} = {\cal B}^{(i_k)}_{i_{k+1}} [{\cal B}^{(i_k)}_{i_{k-1}}]^{-1}
{\cal F} = z_{i_k} e^{(\pm)_{i_k} \frac{\pi}{6} \epsilon}
\label{sturn}
\end{equation}
which was not present in the fundamental representation.

The reason for the appearance of these new signs is the following:
In the spinorial representation, the basic rotations are given by 
\begin{equation}
{\cal B}_{i_{k+1}i_k} = e^{\frac{1}{2} \phi_{i_{k+1}i_k} \epsilon} \, , \quad
{\cal B}_{i_{k-1}i_k} = e^{\frac{1}{2} \phi_{i_{k-1}i_k} \epsilon} \, ,
\end{equation}
where $\phi_{i_{k+1}i_k}$ and $\phi_{i_{k-1}i_k}$ are the angles between 
$(e_{i_k1}, n_{i_{k+1}i_k})$ and $(e_{i_k1}, n_{i_{k-1}i_k})$, respectively.
Therefore, we have
\begin{equation}
{\cal T}_{i_k} = 
e^{\frac{1}{2} (\phi_{i_{k+1} i_k} - \phi_{i_{k-1} i_k} + \pi) \epsilon}
= e^{\frac{1}{2} (\Delta \phi_{i_k} + \pi) \epsilon} \, .
\label{addpi}
\end{equation}
By construction, $\phi_{i_{k+1}i_k}$ and $\phi_{i_{k-1}i_k}$ both lie in the
range $[0,2\pi)$. However, the difference
$\Delta \phi_{i_k} = \phi_{i_{k+1}i_k} - \phi_{i_{k-1}i_k}$ can lie outside
this range. In general, one has $\Delta \phi_{i_k} + \pi = \pm \pi / 3$
modulo $2 \pi$, but $2 \pi$ can be disregarded since
$e^{2 \pi \epsilon} = \mathbbm{1}$. In the spinorial representation, however,
due to the factor $1/2$ one has
$(\Delta \phi_{i_k} + \pi) / 2 = \pm \pi / 6$ modulo $\pi$, and this 
$\pi$ cannot be ignored because $e^{\pi \epsilon} = \pm \mathbbm{1}$.

\begin{figure}
\begin{center}
\includegraphics[height=6cm]{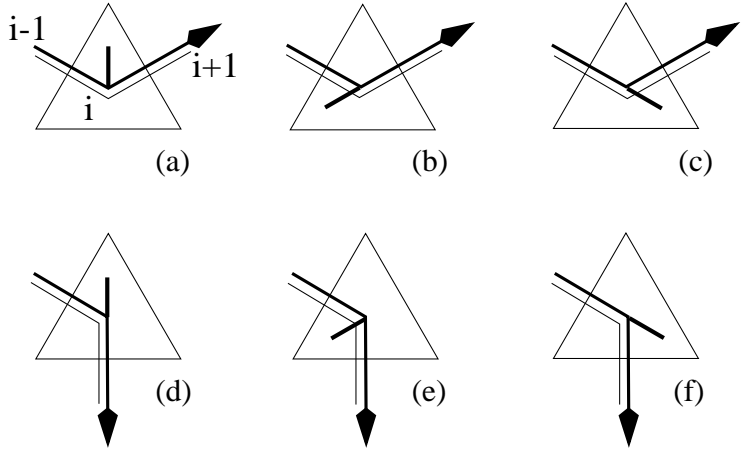}
\caption{\label{Fsix_turns} The six different possibilities for a path to
cross a triangle with a marked $z$-flag, constructed from the two possible
directions of the path (left turn or right turn) and the three possible
directions of the flag. The sign of $\Delta \phi_{i_k}$ is determined by
whether or not the auxiliary line to the right of the path crosses the flag.}
\end{center}
\end{figure}

One has to calculate the exponents in (\ref{addpi}) exactly and to find all
possible values of $\Delta \phi_{i_k}$. There are six different cases, 
collected in fig. \ref{Fsix_turns}.

\begin{figure}
\begin{center}
\includegraphics[height=8cm]{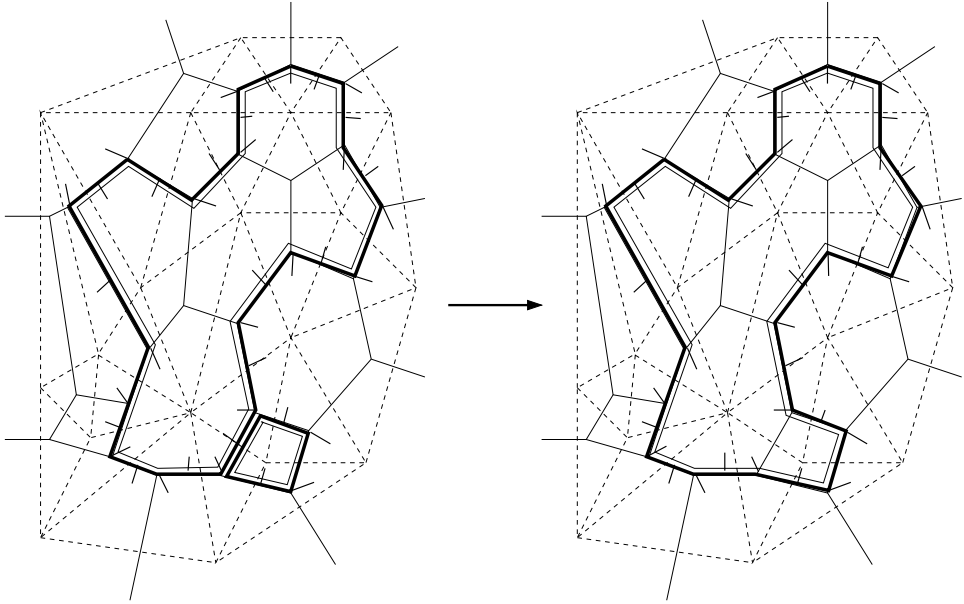}
\caption{\label{Fdefo2} A small deformation of a loop on a triangulated
lattice.}
\end{center}
\end{figure}

The flag in each drawing represents the position of the vector $e_{i_k1}$,
with respect to which the angles are calculated. We call it the $z$-flag.
For example, in the drawing (a) one has $\phi_{i_{k+1}i_k} \in [0, 2\pi/3)$
and $\phi_{i_{k-1}i_k} = \phi_{i_{k+1}i_k} + 4\pi/3$, which yields
 $\Delta \phi_{i_k} = - 4\pi/3$ and thus the rotation matrix~:
\begin{equation}
{\cal T}_{i_k} = e^{\frac{1}{2} (-4 \pi/3 + \pi)}
= e^{-\frac{\pi}{6} \epsilon} \, .
\end{equation}
In the drawing (b) one has $\phi_{i_{k+1}i_k} \in [2\pi/3, 4\pi/4)$ and 
$\phi_{i_{k-1}i_k} = \phi_{i_{k+1}i_k} - 2\pi/3$, so that
$\Delta \phi_{i_k} = 2\pi/3$ and the rotation matrix is
\begin{equation}
{\cal T}_{i_k} = e^{\frac{1}{2} (2\pi/3 + \pi) \epsilon}
= e^{\frac{5\pi}{6} \epsilon} = - e^{-\frac{\pi}{6} \epsilon} \, .
\end{equation}
The results for all six cases (a - f) are given in table \ref{T1}.
\begin{table}
\begin{center}
\begin{tabular}{| c | c | c |}
\hline & & \\
\makebox[1.5cm]{} & \makebox[1cm]{} $\Delta \phi_{i_k}$ \makebox[1cm]{}
& \makebox[0.5cm]{} ${\cal T}_{i_k}
= e^{\frac{1}{2} (\phi_{i_{k+1} i_k} - \phi_{i_{k-1} i_k} + \pi) \,\epsilon}$
\makebox[0.5cm]{} \\
& & \\
\hline & & \\
(a) & $-4\pi/3$ & $+e^{-\pi/6 \, \epsilon}$ \\
(b) & $+2\pi/3$ & $-e^{-\pi/6 \, \epsilon}$ \\
(c) & $+2\pi/3$ & $-e^{-\pi/6 \, \epsilon}$ \\
& & \\
(d) & $-2\pi/3$ & $+e^{+\pi/6 \, \epsilon}$ \\
(e) & $+4\pi/3$ & $-e^{+\pi/6 \, \epsilon}$ \\
(f) & $-2\pi/3$ & $+e^{+\pi/6 \, \epsilon}$ \\
& & \\
\hline 
\end{tabular}
\end{center}
\caption{\label{T1} The difference of angles $\Delta \phi_{i_k}$ and the
turning matrix ${\cal T}_{i_k}$ in the spinorial representation for the
six cases shown in fig. \ref{Fsix_turns}.}
\end{table}
Inserting them into the formula for the loop invariant (\ref{sttt})
one obtains~:
\begin{equation}
{\rm Tr} \, {\cal U}(L) = 
\prod_k s_{i_{k+1}i_k} \, \cdot \, {\rm Tr} \,  \prod_k {\cal T}_{i_k}  
= \prod_k s_{i_{k+1}i_k} z_{i_k} \, \cdot \,
{\rm Tr} \,  \prod_k e^{(\pm)_{i_k} \frac{\pi}{6} \epsilon}
\label{linv}
\end{equation}
where $z_{i_k}$ is the sign of ${\cal T}_{i_k}$. Setting~:
\begin{equation}
S_L = -\prod_k s_{i_{k+1}i_k} z_{i_k} \, ,
\label{Ssz}
\end{equation}
one finds~:
\begin{equation}
{\rm Tr} \, {\cal U}(L) = -S_L \cdot {\rm Tr} \,  
\prod_k e^{(\pm)_{i_k} \frac{\pi}{6} \epsilon} \, .
\label{ut}
\end{equation}
The relation (\ref{Ssz}) between the loop sign $S_L$, the link signs $s$,
and the $z$-signs can be represented graphically in a very intuitive way.
The signs $z_{i_k}$ tell on which side of the path lives the $z$-flag.
If one draws an auxiliary line, as in fig. \ref{Fsix_turns}, that runs
along the right-hand side of the main path, then the sign $z_{i_k}$ can be
determined geometrically by choosing $z_{i_k} = -1$ if the auxiliary line
crosses the $z$-flag and $z_{i_k} = +1$ otherwise. Similarly, one can
introduce a field of flags associated with the oriented links, and
choose the sign $s_{ji} = -1 (+1)$ when the respective $s$-flag is (is not)
crossed when one is going from $i$ to $j$. Because for any given link
the auxiliary path crosses the $s$-flag when going in one direction but not
in the other, this choice leads to $s_{ji} s_{ij} = -1$ as required by
(\ref{ss}). The total sign $S_L$ of the loop $L$ is now given 
by the number of flags ${\rm F}_L$ that are crossed 
by the auxiliary path~:
\begin{equation}
S_L = (-1)^{1+{\rm F_L}} \, .
\label{sloop1}
\end{equation}
As on the regular lattice, one can use the concept of small deformations of
loops to prove some topological theorems for the signs of the loops. The fact
that each elementary loop has $S = +1$ implies that two loops $L$, $L'$ that
can be transformed into each other by a small deformation always have the
same sign, $S_L = S_{L'}$, because a small deformation changes the number
of flags crossed by the loop by an even number (see fig. \ref{Fdefo2}).

Thus, we see that if all
elementary loops on the lattice have positive signs,
all contractible loops have positive signs $S_L = +1$, too. 
Likewise, one can show that all loops belonging to the 
same homotopy class have the same sign. In other words, 
all the topological theorems we found for the regular lattice
hold for the triangulated one as well. The remaining thing is
to check that on a given lattice an assignment of
the link signs $s_{ij}$, ensuring the positivity of 
all elementary loops signs, does always exist. That it is so
for any discretized orientable
2D manifold in \cite{bjk,bjk1}.

\section{The Dirac-Wilson operator\label{secDW}}

We now have all what is needed to construct the fermionic action
(\ref{sh}). We start by casting the formula (\ref{hopping}) for the
hopping operator
\begin{equation}
{\cal H}_{ij} = \frac{1}{2} \left[ 1 - n^{(i)}_{ji} \cdot
\gamma \right] {\cal U}_{ij}
\label{ht}
\end{equation}
into a form that depends on the field of orthogonal frames through the basic
rotations. One can use equation (\ref{ss1}) to decompose the matrix
${\cal U}_{ij}$~:
\begin{equation}
{\cal U}_{ij} = s_{ij} \, 
[{\cal B}^{(i)}_j]^{-1} \epsilon \, {\cal B}^{(j)}_i \quad .
\end{equation}
Likewise, we write the vector $n^{(i)}_{ji}$ in terms of the basic
rotations. By definition, the basic rotations at point $i$ relate the
direction $e_{i1}$ of the frame to the directions of the links between
$i$ and its neighbors $j$~:
\begin{equation}
e_{i1} = [B^{(i)}_j]^{-1} n^{(i)}_{ji} \, .
\end{equation}
In the spinorial representation (\ref{repr}) one can write~: 
\begin{equation}
n^{(i)}_{ji} \cdot \gamma
= [{\cal B}^{(i)}_i]^{-1} \gamma^1 {\cal B}^{(i)}_j \, ,
\end{equation}
where $\gamma^1 = e_{i1} \cdot \gamma$ is the gamma matrix associated with
the first direction of the frame. As mentioned before, the gamma matrices
have the same numerical values $\gamma^1 = \sigma_3$, $\gamma^2 = \sigma_1$
in each frame on the triangulation. Inserting everything into (\ref{ht}) we
eventually obtain~:
\begin{equation}
{\cal H}_{ij} = s_{ij} [{\cal B}^{(i)}_j]^{-1} 
\frac{1}{2} [ 1 - \gamma^1] \epsilon {\cal B}^{(j)}_i \, ,
\label{hf}
\end{equation}
which defines the hopping term in the Dirac-Wilson operator on the
triangulated lattice.

To calculate the basic rotations, one has to find on each triangle the three
angles between $e_{i1}$ and the nearest neighbor vectors $n_{ji}$; denote
them by $\phi^{(i)}_j$. Each is defined in the fundamental range of
the rotation group, $[0, 2\pi)$. Since physical quantities cannot
depend on the choice of the field of frames, we are free to make
the most convenient choice. Hence, we assume that in
each triangle the vector $e_{i1}$ points to one of the
vertices. This implies that the angles $\phi^{(i)}_j$ can take only one of
the three possible values - $\pi/3$, $\pi$ or $5\pi/3$ - which in turn
makes the basic rotation matrices very simple~:
\begin{equation}
{\cal B}^{(i)}_j = e^{\frac{\phi^{(i)}_j}{2} \epsilon}
= \left( \begin{array}{rr} c^{(i)}_j & s^{(i)}_j \\ 
                          -s^{(i)}_j & c^{(i)}_j \end{array} \right) \, ,
\end{equation}
where
\begin{equation}
c^{(i)}_j \equiv \cos \frac{\phi_j^{(i)}}{2} = 
\frac{\sqrt{3}}{2},0,-\frac{\sqrt{3}}{2} \quad , \qquad
s^{(i)}_j \equiv \sin \frac{\phi_j^{(i)}}{2} =
\frac{1}{2}, 1 , \frac{1}{2} 
\end{equation}
for $\phi_j^{(i)} = \pi/3, \pi, 5\pi/3$, respectively.
Inserting this explicit form of the basic rotations into (\ref{hf}) leads to
an extremely simple formula because $(1 - \gamma^1)/2$ is a
projection matrix, which with our choice of $\gamma^1$ has only one
non-vanishing element. Hence~:
\begin{equation}
{\cal H}_{ij} = s_{ij} 
\left(\begin{array}{rr} s^{(i)}_j c^{(j)}_i & s^{(i)}_j s^{(j)}_i \\ 
                       -c^{(i)}_j c^{(j)}_i &-c^{(i)}_j s^{(j)}_i
\end{array}\right) \, .
\label{H}
\end{equation}
In this form the Dirac-Wilson operator is easy to implement. For each pair
of neighboring triangles $j$ and $i$ we first find the sign $s_{ji}$ and
the angles between the $z$-flag and the dual link $ji$ and calculate the
appropriate trigonometric functions. For example, assuming $s_{ij} = 1$
for the link $ji$ in fig. \ref{FBrot} we have $\phi^{(j)}_i = {5\pi}/3$,
$\phi^{(i)}_j = \pi$. Hence~:
\begin{equation}
{\cal H}_{ij} = \left(\begin{array}{rr} -\frac{\sqrt{3}}{2}  &  \frac{1}{2} \\ 
                      0 & 0 \end{array}\right) \, .
\end{equation}
The Dirac-Wilson operator is built from blocks like the above one,  
for each pair of indices representing neighboring triangles, and 
from $2 \times 2$ unit matrices for each pair of identical 
indices. Defining the adjacency matrix for triangles as~:
\begin{equation}
{\cal A}_{ij} = \left\{ \begin{array}{ll} 1 & 
{\rm if} \ i \ {\rm and} \ j \ {\rm are \ neighbors} \\
0 & {\rm otherwise} \end{array} \right.
\end{equation}
one can write the Dirac-Wilson operator as~:
\begin{equation}
{\cal D}_{ij} = -K {\cal A}_{ij} {\cal H}_{ij}
+ \delta_{ij} \mathbbm{1} \, .
\label{D}
\end{equation}
What are the properties of the Dirac-Wilson operator in this form? 
Consider the charge conjugation transformation~:
\begin{equation}
\psi \to \psi_c = C \bar{\psi}^T \, , \quad 
\bar{\psi} \to \bar{\psi}_c = - \psi^T C^{-1} \, ,
\label{cc}
\end{equation}
where the matrix $C$ is unitary and fulfills the requirements~:
\begin{equation}
C^{-1} \gamma^T C = -\gamma \, , \quad 
C^T = -C \, .
\end{equation}
One can check that the hopping operator (\ref{hf}) transforms as~:
\begin{equation}
C {\cal H}^T_{ij} C^{-1} = {\cal H}_{ji} \, .
\label{chc}
\end{equation}
In two dimensions we can choose the standard antisymmetric matrix
$\varepsilon$ as the charge conjugation matrix, $C=\varepsilon$.
It is convenient to use two different versions of $\varepsilon$, one with
lower indices $\varepsilon_{\alpha\beta}$ and one with upper indices
$\varepsilon^{\alpha\beta}$, but with the same numerical values~:
\begin{equation}
\varepsilon_{12} = \varepsilon^{12} = 1 \, , \quad
\varepsilon_{\alpha\gamma} \varepsilon^{\gamma\beta}
= -\delta_\alpha^\beta \, .
\end{equation}
One can treat $\varepsilon$ as a simplectic form to raise or lower
the spinorial indices~:
\begin{equation}
(\psi_c)_\alpha = \varepsilon_{\alpha\beta} \psi^\beta 
\, , \quad
(\psi_c)^\alpha = \psi_\beta \varepsilon^{\beta\alpha} \, .
\label{ccpsi}
\end{equation}
We recall that in the explicit index notation, the components of the spinor
$\bar{\psi}$ are denoted by $\psi^\alpha$ and those of $\psi$ by $\psi_\alpha$.
Furthermore, in this notation one can write~:
\begin{equation}
{\cal D}_{ij}^{\alpha\beta} = \varepsilon^{\alpha\gamma} 
[{\cal D}_{ij}]_\gamma^\beta \, .
\end{equation}
In the implicit index notation one has to distinguish between different
cases, namely ${\cal D}$ for mixed indices, $\varepsilon {\cal D}$ for only
upper indices, and ${\cal D} \varepsilon$ for only lower indices, by
displaying explicitly the action of $\varepsilon$.

The fact that the hopping operator is constructed from a projector implies in
particular, that~:
\begin{equation}
{\cal H}_{ij} {\cal H}_{ji} = 0 \, , \quad
{\cal H}_{ij} {\cal U}_{ji} {\cal H}_{ij} = {\cal H}_{ij} \, .
\label{noback}
\end{equation}
The consequence of the transformation law (\ref{chc}) is that~:
\begin{equation}
\varepsilon {\cal H}_{ij} \varepsilon = -{\cal H}_{ji}^T
\label{cch}
\end{equation}
and, furthermore, that~:
\begin{equation}
\left( \varepsilon {\cal D}_{ij} \right)^T = -\varepsilon {\cal D}_{ji} \, .
\label{ccd}
\end{equation}
In index-explicit notation, this last equation reads~:
\begin{equation}
{\cal D}_{ij}^{\alpha\beta} = - {\cal D}_{ji}^{\beta\alpha}  \, ,
\label{antisym}
\end{equation}
which means that the matrix ${\cal D}_{ij}^{\alpha\beta}$ is antisymmetric
in the double indices $I = (i\alpha)$ and $J = (j\beta)$~:
${\cal D}_{IJ} = -{\cal D}_{JI}$.

\section{Second-quantized theory\label{secSQ}}

Quantum field theory of free Dirac fermions in a curved geometrical
background represented by a triangulation $T$ is defined by the
partition function~:
\begin{equation}
Z_T(K) = \int \prod_i {\rm d}^2 \psi_i {\rm d}^2 \bar{\psi}_i 
e^{-\psi_i^\alpha [{\cal D}_{ij}]_\alpha^\beta \psi_{i\beta}}
= | {\cal D} | \, .
\label{ZT}
\end{equation}
The propagator is~:
\begin{equation}
\left\langle \psi_{n\nu} \psi_m^\mu \right\rangle
= \frac{1}{Z_T(K)} \int \prod_i {\rm d}^2 \psi_i
{\rm d}^2 \bar{\psi}_i \, \psi_{n\nu} \psi_m^\mu \,
e^{-\psi_i^\alpha [{\cal D}_{ij}]_\alpha^\beta \psi_{i\beta}}
= [{\cal D}^{-1}_{nm}]_\nu^\mu \, . 
\end{equation}
It transforms under a local change of frames $e_i \to e'_i = R_i e_i$ as
follows~:
\begin{equation}
\left\langle \psi_{n\nu} \psi_m^\mu \right\rangle 
\ \to \ \left\langle \psi'_{n\nu} {\psi'}_m^\mu \right\rangle =
[{\cal R}_n]_\nu^\alpha [{\cal R}^{-1}_m]^\mu_\beta
\left\langle \psi_{n\alpha} \psi_m^\beta \right\rangle \, .
\end{equation} 
Let us further explore the consequences of the symmetry with respect to the
charge conjugation that is encoded in the transformation law (\ref{chc}).
Introduce two families of Majorana fermions~:
\begin{eqnarray}
\phi_1 = \frac{1}{2}(\psi_c+\psi) \, , & &
\bar{\phi}_1 = \frac{1}{2}(\bar{\psi}_c+\bar{\psi}) \, , \nonumber \\
\phi_2 = \frac{1}{2i}(\psi_c-\psi) \, , & &
\bar{\phi}_2 = \frac{-1}{2i}(\bar{\psi}_c-\bar{\psi}) \ .
\end{eqnarray}
They are charge self-conjugate~:
$\phi_{1 c} = \phi_1$ and $\phi_{2 c} = \phi_2$. 
This means that the components of $\phi_1$
are not independent, likewise for $\phi_2$.
The components are related~:
\begin{equation}
\phi^\alpha = \phi_\beta \varepsilon^{\beta\alpha} .
\end{equation}
as can be seen from (\ref{ccpsi}). 
We skipped the family index $1,2$ in the last formula.

It is convenient to express the Dirac-Wilson action in terms of the
Majorana families $\phi_1$ and $\phi_2$. Indeed, using equation (\ref{chc})
one finds that the two families decouple~:
\begin{equation}
S (\bar{\psi}, \psi) = \frac{1}{2} \sum_i \bar{\psi}_i \psi_i
- K \sum_{\langle ij \rangle} \bar{\psi}_i {\cal H}_{ij} \psi_j
= {\cal S} (\phi_1) + {\cal S} (\phi_2) 
\end{equation}
where
\begin{equation}
{\cal S}(\phi) = \frac{1}{2} \sum_i \bar{\phi}_{i} \phi_{i} 
- K \sum_{\langle ij \rangle} \bar{\phi}_{i} {\cal H}_{ij} \phi_{j} \, .
\end{equation}
The two actions $S (\bar{\psi}, \psi)$ and ${\cal S} (\phi)$ appear identical
to each other, but they differ in the number of degrees of freedom; in the
latter case, $\bar{\phi}$ is uniquely determined by $\phi$. By changing
the variables in the integration measure of (\ref{ZT}) one can rewrite the
partition function as a product of two identical factors~:
\begin{equation}
Z_T (K) = \int \prod_i {\rm d}^2 \phi_{1i} {\rm d}^2 \phi_{2i} \,
e^{-S(\phi_1) - S(\phi_2)} = \left[ {\cal Z}_T(K) \right]^2
\end{equation}
where ${\cal Z}_T(K)$ is the partition function for a single Majorana family~:
\begin{eqnarray}
{\cal Z}_T (K) & = & \int \prod_i {\rm d}^2 \phi_i \,
e^{-\frac{1}{2} \sum_i \bar{\phi}_i \phi_i + K \sum_{\langle ij \rangle}
\bar{\phi}_i {\cal H}_{ij} \phi_j} \nonumber \\
& = & \int \prod_i {\rm d}^2 \phi_i e^{-\phi_{i\alpha}
{\cal D}_{ij}^{\alpha\beta} \phi_{i\beta}}
= {\rm Pfaff} [\varepsilon {\cal D}] \, .
\label{Pfaffian}
\end{eqnarray}
Here, $\varepsilon {\cal D}$ is the antisymmetric matrix (\ref{antisym}),
which implies that the square of the Pfaffian is equal to the determinant of
$\varepsilon {\cal D}$, which is in turn equal to the determinant of
${\cal D}$. We can calculate the partition function 
for the Majorana fermions using the hopping parameter
expansion. This leads to a geometrical interpretation of the model, as 
will be seen in the next section.

\section{Fermionic loops\label{secFL}}

To find the hopping parameter expansion of ${\cal Z}_T (K)$
let us first split the integrand into two parts~:
\begin{equation}
{\cal Z}_T(K) = \int \prod_i \left( {\rm d}^2 \phi_i \,
e^{-\frac{1}{2} \bar{\phi}_i \phi_i}  \right) \ \prod_{\langle ij \rangle }
\left( 1 + K \bar{\phi}_i {\cal H}_{ij} \phi_j \right)
\label{hexp0}
\end{equation}
The first part is a product of independent one-point integrations with an
exponential measure, whereas the second is a product over all oriented links
that connect neighboring points. Since we know from equations (\ref{chc})
and (\ref{cch}) that for Majorana fermions~:
\begin{equation}
\bar{\phi}_j {\cal H}_{ji} \phi_i = \bar{\phi}_i {\cal H}_{ij} \phi_j \, ,
\end{equation}
it is convenient to rewrite the product in (\ref{hexp0}) as a product over
non-oriented links $(ij)$~:
\begin{equation}
{\cal Z}_T(K) = \int \prod_i \left( {\rm d}^2 \phi_i \,
e^{-\frac{1}{2} \bar{\phi}_i \phi_i} \right) \
\prod_{(ij)} \left( 1 + 2 K \bar{\phi}_i {\cal H}_{ij} \phi_j \right) \, .
\label{hexp}
\end{equation}
To do this, we have to require that terms like
$\bar{\phi}_i {\cal H}_{ij} \phi_j \bar{\phi_j} {\cal H}_{ji} \phi_i$
do not occur in the expansion. Actually, they vanish because
of (\ref{noback}).

The only non-vanishing integrals relevant to our problem are~:
\begin{equation}
\int {\rm d}^2 \phi \ e^{-\frac{1}{2} \bar{\phi} \phi} \cdot 1 = 1
\label{unity}
\end{equation}
and
\begin{equation}
\int {\rm d}^2 \phi \ e^{-\frac{1}{2} \bar{\phi} \phi}
\cdot (\phi \cdot \bar{\phi}) = \mathbbm{1} \, . 
\label{delta}
\end{equation}
These rules are used to calculate the integral of each term in the
expansion~:
\begin{eqnarray}
\prod_{(ij)} ( 1 + 2K \bar{\phi}_i {\cal H}_{ij} \phi_j )
& = & 1 + 2K \sum_{(ij)} \bar{\phi}_i {\cal H}_{ij} \phi_j \nonumber \\
& & + (2K)^2 \sum_{(ij),(kl)} \bar{\phi}_i {\cal H}_{ij} \phi_j 
\cdot \bar{\phi}_k {\cal H}_{kl} \phi_l + \dots
\label{expansion}
\end{eqnarray}
Consider the quadratic term on the right hand side. If $j = k$ then, 
according to (\ref{delta}), the integration over $\phi_j$ yields~:
\begin{equation}
\sum_{(ij),(jl)} \bar{\phi}_i {\cal H}_{ij} \phi_j 
\cdot \bar{\phi}_j {\cal H}_{jl} \phi_l = \sum_{(ij),(jl)}  
\bar{\phi}_i ({\cal H}_{ij} {\cal H}_{jl}) \phi_l \, .
\end{equation}
Otherwise, if $j \ne k$, the integral vanishes. In general, one observes
that the contribution of a term in the expansion (\ref{expansion}) is non-vanishing
only when all neighboring fields $\phi_j \cdot \phi_k$ belong to
the same point. Integration of these terms over all fields gives~:
\begin{equation}
\bar{\phi}_{j_1} {\cal H}_{j_1j_2} {\cal H}_{j_2j_3} \cdots 
{\cal H}_{j_{n-1}j_n} \phi_{j_n} \, ,
\end{equation}
where all $j_i$ in the chain are different. For the final integration to
yield something non-vanishing one must have $j_1 = j_n$. 
Finally~:
\begin{equation}
C (L) = - {\rm Tr} \, {\cal H}_{j_1j_2} {\cal H}_{j_2j_3} \cdots 
{\cal H}_{j_{n-1}j_1} \, .
\label{lc}
\end{equation}
This contribution can be graphically represented by a closed loop
$L = (j_1, j_2, \dots, j_{n-1}, j_1)$ of length $n$. On the other hand,
integration over a field $\phi_k$ associated with a vertex $k$ that does not
lie on any loop contributes a factor of 1 (\ref{unity}).

In summary, all terms of the expansion that survive the integration
(\ref{hexp}) can be represented graphically as diagrams consisting of closed
loops. These loops do not back-track or touch each other. A configuration
consisting of $l$ loops $L_1, L_2, \dots, L_l$ with total length
$n = n_1 + \dots + n_l$ contributes a term
\begin{equation}
(2K)^n C (L_1) C (L_2)\dots C (L_l)
\label{mlc}
\end{equation}
to the partition function.

One can calculate the contribution $C (L)$ of a single loop $L$ in a way
similar to that used to obtain the loop invariant (\ref{sttt}), $i.\,e.$
by extracting the total sign of the loop (\ref{Ssz}) and expressing the
remaining product in terms of turns at the vertices (\ref{sturn}). The result
is~:
\begin{equation}
C(L) = - {\rm Tr} \, \prod_k {\cal H}_{i_{k+1},i_k} 
= S_L \cdot {\rm Tr} \, 
 \prod_k {\cal T}_{i_k} \frac{1}{2} (1 - \gamma^1)
\end{equation}
The difference between this expression and the one for the loop invariant
(\ref{linv}) is that now in addition to the turn matrix a
projection operator appears in the product. Inserting the explicit form
of the turn matrix $T_i = e^{\pm \epsilon \pi/6}$ and of the projector
$(1 - \gamma^1)/2 = (1 - \sigma_3)/2$, one obtains~:
\begin{equation}
C(L) = S_L \left( \frac{\sqrt{3}}{2} \right)^n \, .
\label{cln}
\end{equation}
This is again similar to the result found for the loop invariant (\ref{ut}), but
with two differences. First, one now has $S_L$ instead of $- S_L$.
Second, in the calculation of the loop invariant the turn
angles enter the result with a sign $\pm$ depending on whether the path turns
left or right, whereas here the projector leaves only the cosines of the
rotation matrix, which depend on the absolute value of the turn angle. Thus,
each turn contributes a factor $+\sqrt{3}/2$ independently of its direction.
Since a loop makes a turn at each vertex, the number of turns in a loop is
simply equal to the loop length, which gives (\ref{cln}).

\begin{figure}
\begin{center}
\includegraphics[height=6cm]{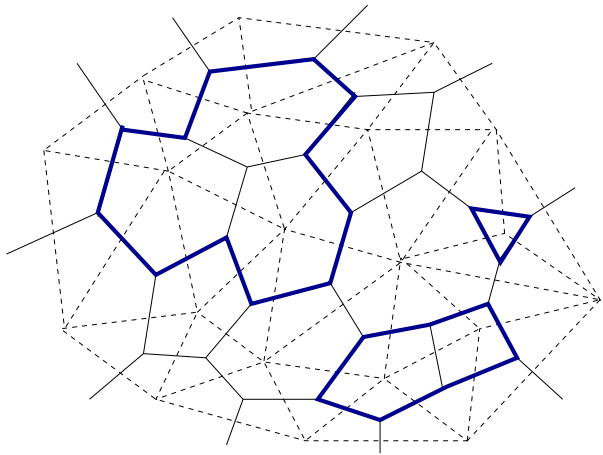}
\caption{\label{Ffloops} A configuration of fermionic loops.}
\end{center}
\end{figure}

Inserting this result into (\ref{mlc}), one finds that the contribution of a
loop configuration of total length $n$ is~:
\begin{equation}
S_{total} \cdot \left(\sqrt{3} K \right)^n \, ,
\label{Ksqrt3}
\end{equation}
where
\begin{equation}
S_{total} = \prod_L S_L \, .
\end{equation}
On a lattice with spherical topology all loops $L$ have a positive sign
$S_L = 1$ and therefore $S_{total} = 1$ for each loop configuration. 

On a torus, the sign of the contribution depends on the spin structure. 
Assuming periodic boundary conditions in both directions $(++)$, all
loops from any non-trivial homology classes, contractible or not,
have $S_L = 1$, and again $S_{total} = 1$ for any loop
configuration. The standard notation is used here:
the spin structure is referred to by the signs of independent classes 
of non-contractible loops. On the torus there are two classes and
therefore four possibilities $(ss')$, with $s,s' = \pm$. Plus/minus corresponds to
periodic/antiperiodic boundary condition for spinors transported
along loops in this class. With anti-periodic boundary conditions
in any direction -  $(+-)$, $(-+)$, or $(--)$ - any non-contractible
loop circling the lattice in this direction has a
negative sign $S_L = -1$. Thus, all of these three cases can produce unwanted
configurations with a negative contribution to the partition function.
More generally, any configuration that has an odd number of non-contractible
loops circling the lattice in an anti-periodic direction has a negative
total sign $S_{total} = -1$.

Yet another possible choice of boundary conditions imposes summation over all
spin structures -  $(++)$, $(+-)$, $(-+)$, and $(--)$ - in the partition
function. This operation is called GSO projection, and in many cases seems to be the
most physical choice. Negative contributions are not a problem in this case~:
 a configuration with an odd numbers of loops in one of the
non-trivial homotopy classes, say in the first class of non-contractible loops,
has $S_{total} = 1$ for $(++)$ and $(+-)$, but $S_{total} = -1$ for $(-+)$ and
$(--)$. The summation over all cases yields zero.
More generally all `bad' contributions to the partition function
cancel out in the GSO projection.

\begin{figure}
\begin{center}
\includegraphics[height=4cm]{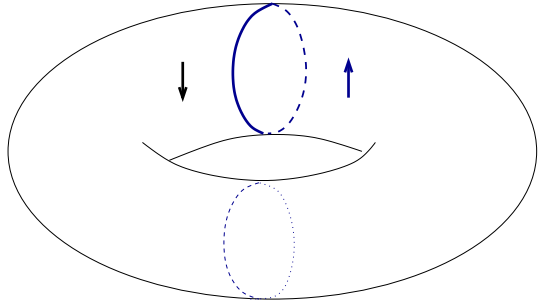}
\caption{\label{Ftdwalls} Domain walls versus loops on a torus. 
A non-contractible loop on a torus, like for instance
the upper curve in the figure, cannot be a part of the domain wall
configuration of Ising spins unless
there is a partner curve in the same class of loops in this
configuration, like for example the lower one. 
In general, domain-wall configurations of 2D Ising model
have an even number of loops in each non-trivial class of 
non-contractible loops.}
\end{center}
\end{figure}

Configurations with an odd number of non-contractible
loops in at least one direction cannot correspond to Ising model domain wall
configurations, because only an even number of these domain walls is
crossed when one is performing a round trip on the lattice (see
fig. \ref{Ftdwalls}). From this point of view, the loop cancellation in
GSO projection is very physical. Before discussing this point in more detail,
a more careful look at the properties of the loop signs is needed.

\section{The GSO projection\label{secGSO}}

As discussed in the preceding sections, the 
global properties of the Dirac-Wilson operator on a
two-dimensional compact manifold are closely related to the signs of the
fermionic loops. Self-consistency requires a positive sign for all
elementary fermionic loops, and this in turn implies a positive sign for all
contractible loops. Non-contractible loops, on the other hand, are not subject
to this restriction. In fact, it is the ensemble of signs of all independent 
non-contractible loops that defines the spin structure of the manifold.

In this section, it will be shown that the sign of any loop on the lattice
is uniquely determined by the signs of a minimal number of independent 
non-contractible loops. Stated differently: 
the signs of all loops on the manifold 
are completely encoded in the manifold's spin structure.

So far, we discussed the loops without self-crossings only, for the simple
reason that on a triangulation no other loops occur in the hopping parameter
expansion of the Majorana-Dirac-Wilson fermions. On the other hand, we already
encountered self-crossing implicitly in the calculation of the invariants
${\rm Tr} \, {\cal U} (C)$ (\ref{sttt}), since they can be defined on loops 
of any kind, including the self-crossing ones.\footnote{It is convenient to think of
a self-crossing on a lattice not as a meeting at exactly one vertex, but
rather as a sort of smeared overlapping that may occupy one or more links
of the lattice. In particular, on a lattice with only vertices of order three,
there are no exact one-vertex self-crossings; the most localized ones still
occupy at least one link.}
 
\begin{figure}
\begin{center}
\includegraphics[height=3cm]{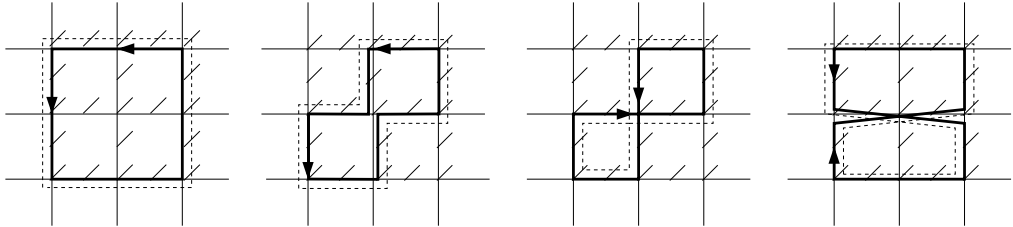}
\caption{\label{Fclsigns}Examples of contractible loops with and without
self-crossings on a square lattice. The numbers of crossed flags and
self-crossings are $F_1 = 9$, $C_1=0$ for the first example; $F_2 = 9$,
$C_2=0$ for the second (the flag in the center 
of the figure is crossed twice!)~;
 $F_3 = 8$, $C_3=1$ for the third; and
$F_3 = 12$, $C_3=1$ for the last one. The result is a positive sign
in all cases. Note that in the last three examples, the flag at the vertex
in the center, where four links of the loop meet, is crossed an even number
of times by the auxiliary line.}
\end{center}
\end{figure}

For this reason, and also for the sake of completeness, we shall now discuss the
signs and topological properties of loops in a general context, and
restrict them to self-avoiding loops only when necessary. We  require
the sign of a loop to be a property of its homotopy, which means that we
have to modify the definition of the sign (\ref{sloop1}) to~:
\begin{equation}
S_L = (-1)^{1 + {\rm F_L} + {\rm C_{LL}}} \, ,
\label{sloop2}
\end{equation}
where ${\rm C}_{LL}$ is the number of self-crossings of the loop $L$. 
Of course, for any contractible loop this must still result in a positive sign,
independently of the number of self-crossings. A few examples of contractible
loops with various numbers of self-crossings are shown in fig. \ref{Fclsigns}.
It is easy to verify that the auxiliary line running along the right hand side
of the loop crosses an odd number of flags in the first two cases and an even
number of flags in the last two.

\begin{figure}
\begin{center}
\includegraphics[height=6cm]{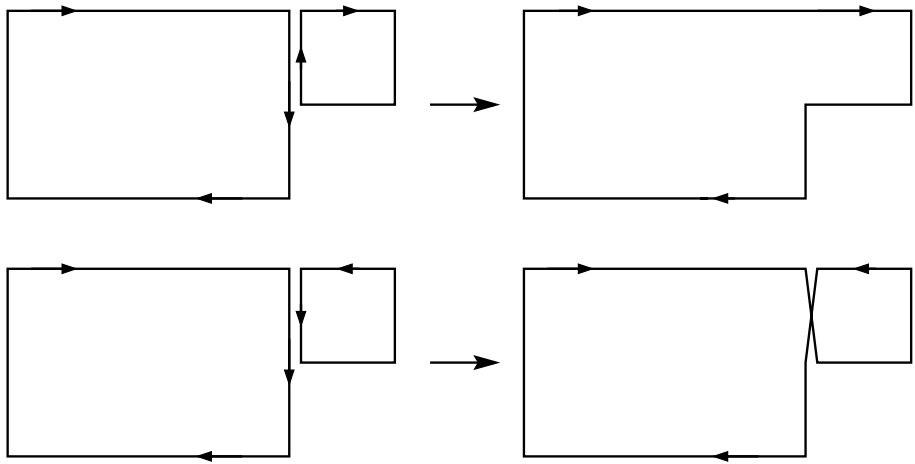}
\caption{\label{Fdefo3}Even and odd versions of a 'small deformation'. In
the upper figure, the two loops contain a common link of opposite orientation,
causing the two versions of the link to `cancel out' in the resulting deformed
loop. In the lower figure, the common link has the same orientation in both
loops, causing it to appear twice in the deformed loop and thus introducing a
self-crossing.}
\end{center}
\end{figure}

Let us now return to the operation that we called a 'small deformation'. So far,
we have considered only deformations that do not induce self-crossings (see
for example fig. \ref{Fdefo2}). These deformations will be called {\em even}. It
is convenient to introduce also an {\em odd} version of a small deformation, where
an elementary plaquette is again used to deform the loop, but with a
a self-crossing like in fig. \ref{Fdefo3}. The two kinds of
small deformations differ by the orientation of the plaquette that is used to
deform the loop. Neither kind changes the overall sign of the deformed loop.

One can introduce equivalence classes of loops that can be obtained from each
other by a sequence of small
deformations. A class of loops equivalent to a loop $A$ will be denoted by
$[A]$. Inside this class, $[A]_{even}$ denotes the sub-class of loops
that can be obtained from $[A]$ by a sequence of an even number of 
small deformations.

\begin{figure}
\begin{center}
\psfrag{A}{{\footnotesize $A$}}
\psfrag{B}{{\footnotesize $B$}}
\psfrag{A B}{{\footnotesize $A\cdot B$}}
\includegraphics[height=3.1cm]{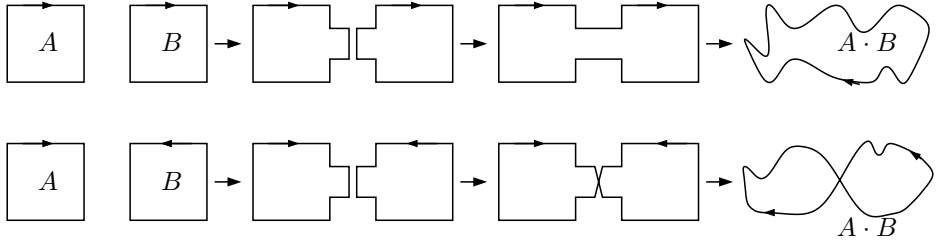}
\caption{\label{Flmerging} The loop merging operation. Two loops $A \in [A]$
and $B \in [B]$ are smoothly deformed until they share a link. They are then
joined either by erasing the link or by a self-crossing, depending on the
link's relative orientation in both loops. The resulting loop belongs to
a new class of loops $[A \cdot B]$.}
\end{center}
\end{figure}

Let us define the {\em loop merging} operation, that acts on a set
of equivalence classes of loops. Take two loops $A \in [A]$ and $B \in [B]$,
and deform both of them smoothly until they have a common link. If this common
link has an opposite orientation in both loops, erase it and form a loop out
of the remaining links. Otherwise, leave the link as it is and join $A$ and
$B$ by a self-crossing (see fig. \ref{Flmerging}). The resulting loop belongs, by
definition, to a new equivalence class of loops $[A \cdot B]$.

The product of loop classes defined in this way has a unity element in the
class of contractible loops $[E]$, for which 
$[A \cdot E] = [E \cdot A] = [A]$. Counting the number of crossed flags
before and after the loop merging (steps 2 and 3 in fig. \ref{Flmerging}),
one finds~:
\begin{equation}
F_{[A \cdot B]} \stackrel{mod \ 2}{=} F_{[A]} + F_{[B]} + 1 + C_{AB} \, ,
\end{equation}
where $C_{AB}$ is the number of crossings of the loops $A$ and $B$.
The equation implies the law of sign composition~:
\begin{equation}
S_{[A\cdot B]} = S_{[A]} \, S_{[B]} \, .
\label{scomp}
\end{equation}
Indeed, in the upper drawing in fig. \ref{Flmerging} the loop merging does not
introduce any additional self-crossing, $C_{AB} = 0$, and the number of
crossed flags changes by $1$ modulo $2$, whereas in the lower figure 
one additional self-crossing appears, $C_{AB} = 1$, and the number of crossed
flags changes by $0$ modulo $2$, $i.\,e.$ it remaines unaltered. The factors
coming from the flag count and from the number of additional self-crossings
compensate each other, and the above simple composition law (\ref{scomp}) follows.

\begin{figure}
\begin{center}
\psfrag{h1}{{\scriptsize $H_1$}}
\psfrag{h2}{{\scriptsize $H_2$}}
\psfrag{h3}{{\scriptsize $H_3$}}
\psfrag{h4}{{\scriptsize $H_4$}}
\psfrag{h1q}{{\scriptsize $H_{2g-1}$}}
\psfrag{h2q}{{\scriptsize $H_{2g}$}}
\includegraphics[height=2.3cm]{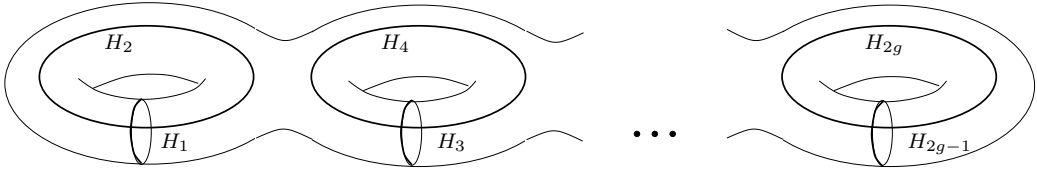}
\caption{\label{FHs} Independent classes of non-contractible loops on a 2d
manifold with genus $g$.}
\end{center}
\end{figure}

Thus,  the set of equivalence classes of loops forms a group with
respect to the loop merging operation. On a two-dimensional compact manifold, 
this group contains a minimal set of independent classes of non-contractible
loops $[H_i]$, $i = 1, \dots, 2g$, where $g$ is the genus of the manifold
(see fig. \ref{FHs}).

This minimal set has the nice feature that all other classes can be created
from $[E]$, $[H_i]$, and their inverses $[H_i]^{-1}$ by use of the loop
merging operation. In other words, one can decompose any loop in terms of
$[E]$ and $[H_i]$, and then use equation (\ref{scomp}) to calculate the sign
of this loop as a product of signs of the $H_i$.

Let us illustrate this with a few examples. For simplicity, denote the signs
of the classes in the minimal set with $S_i \equiv S_{[H_i]} = S_{[H_i]^{-1}}$.

\begin{figure}
\begin{center}
\includegraphics[height=5cm]{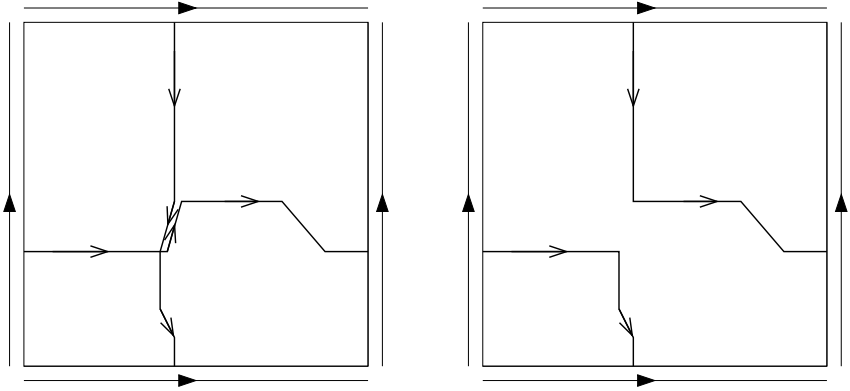}
\caption{\label{FLMa}Loop merging on a torus. Take two loops from the classes
$[H_1]$ and $[H_2]$ and smoothly deform them until they share a link, then
merge them. The resulting loop $[H_1 \cdot H_2]$ circles the torus in both
directions $[H_1]$ and $[H_2]$ simultaneously. Its sign is the product
$S_1 \cdot S_2$.}
\end{center}
\end{figure}

Consider first a loop which goes around a torus in two distinct homotopy
directions simultaneously. Such a loop can be obtained by loop
merging of the classes $[H_1]$ and $[H_2]$, as shown in fig. \ref{FLMa}. 
Note that the loops shown in the figure do not self-cross; nor does the
resulting loop $H_1 \cdot H_2$. This might seem surprising at first, given
that the loop merging itself introduces a crossing, $C_{H_1 H_2} = 1$. But 
indeed one can see that the original loops, even if not
self-crossing, do cross each other. In general, any loop from $[H_1]$ always
crosses any loop from $[H_2]$ an odd number of times. In the resulting merged
loop, these crossings become self-crossings, so that the product has an even
number of self-crossings overall. This in turn means it can be deformed by a
sequence of an even number of small deformations to a non-self-crossing loop.

As a general definition, one can state that a class $[A]$ crosses a class $[B]$
if the number of crossings between any two representatives $A$ and $B$ is odd.
By this definition, the classes $[H_1]$ and $[H_2]$ cross each other, as do
any two of the classes $[H_{2i-1}]$ and $[H_{2i}]$ shown in fig. \ref{FHs}.
This concept will be useful in a while in the context of the
$GSO$ projection.

\begin{figure}
\begin{center}
\psfrag{h1}{{\footnotesize $H_1$}}
\psfrag{h2}{{\footnotesize $H_3$}}
\psfrag{h3}{{\footnotesize $C$}}
\includegraphics[height=6cm]{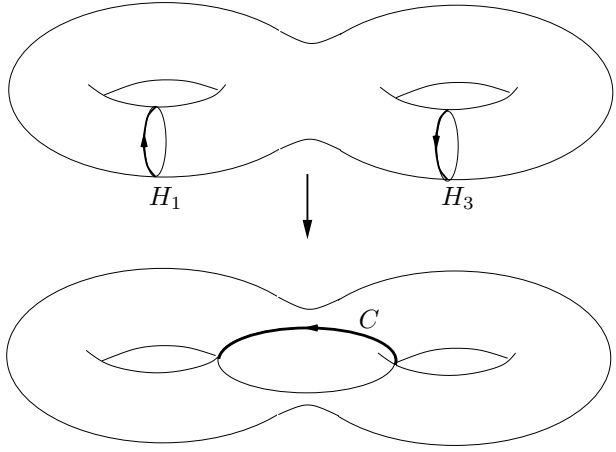}
\caption{\label{FLMb}Loop merging on a double torus. The sign of the loop in
the lower figure can be calculated by observing that it can be created by a
merging of the loops drawn in the upper figure. This can be done by first
deforming the two loops until they have a common link, then erasing this
link, and smoothly deforming the remaining loop.}
\end{center}
\end{figure}

Another example of loop merging is shown in fig. \ref{FLMb}. The loop $C$ in
the lower drawing is obtained by merging $H_1$ and $H_3$, so its sign can be
calculated as the product $S_C = S_{H_1} S_{H_3}$.

Let us apply the sign composition law to the calculation of the partition
function of Majorana fermions (\ref{hexp}). In the hopping expansion, one
generates non-self-crossing loops only. Denote the number of loops from
a given class $S_{[C]}$ on a configuration by $N_{[C]}$. Then the total sign
of this configuration can be written as~:
\begin{equation}
S_{total} = \prod_{[C]} \left( S_{[C]} \right)^{N_{[C]}} \, .
\label{sct}
\end{equation}
After GSO projection~:
\begin{equation}
{\cal Z}^{GSO} = \frac{1}{2^{2g}}
\sum_{\{ (\pm)_{1, \dots, 2g} \}}
{\cal Z}^{\left( (\pm)_1, (\pm)_2, \dots, (\pm)_{2g} \right)} \, ,
\end{equation}
the total contribution of the configuration is proportional to
\begin{equation}
{\cal W}^{GSO} \sim
\prod_{[C]} \frac{1}{2} \left( 1 + (-1)^{N_{[C]}} \right) \, .
\end{equation}
To see this, note first of all that the sum over the signs $S_i$ of the
classes $[H_i]$ can be replaced by a sum over the signs $S_{[C]}$ of the
classes $[C]$ present in the configuration, since all these loops do
not cross and are independent from each other. Summing over all signs
$S_{[C]}$ means that each loop of each non-trivial class $[C]$ occurs an
equal number of times with plus and minus signs, which eventually leads to
the last formula. In a sense, the action of the $GSO$ projection factorizes
into a product of independent actions for the loops of each non-trivial class
on the configuration.

The last equation also tells us that all configurations with an odd number
of loops from any non-trivial class have a vanishing contribution to the
$GSO$ projection. Physically, this means that the projection removes all
loop configurations which cannot represent domain wall configurations.

\section{Topology of the Ising model\label{secTI}}

We shall consider now the Ising model with nearest neighbor interactions, 
focusing on the issue of the exactness of the duality
transformation between the model defined on a triangulation and 
on its dual graph, respectively, and emphasizing the
topological aspect of the duality. Furthermore, the relation between the
Ising and the fermionic model will be discussed.

To distinguish between a triangulation and its dual, we attach a star to
symbols referring to the triangulation, while the unstared symbols refer
to the dual lattice.

With this convention, the partition function of the Ising spins living on
the triangulation reads~:
\begin{equation}
\Omega_{T_*} (\beta_*) = \Omega_{T_*}^{(++)}(\beta_*) = 
\sum_{\{ \sigma_{i_*} \}}
e^{\beta_* \sum_{(i_* j_*)} \sigma_{i_*} \sigma_{j_*}} \, ,
\label{omega1}
\end{equation}
where $\sigma_{i_*} = \pm 1$ are spin variables located 
at the vertices $i_*$ of the triangulation. As we shall
see later discussing boundary conditions 
for the Ising model, the partition function (\ref{omega1})
corresponds to the partition function with
the spin structure $(++)$. Therefore we additionally denoted it 
by $\Omega_{T_*}^{(++)}(\beta_*)$ in the last equation.

\begin{figure}
\begin{center}
\includegraphics[height=6cm]{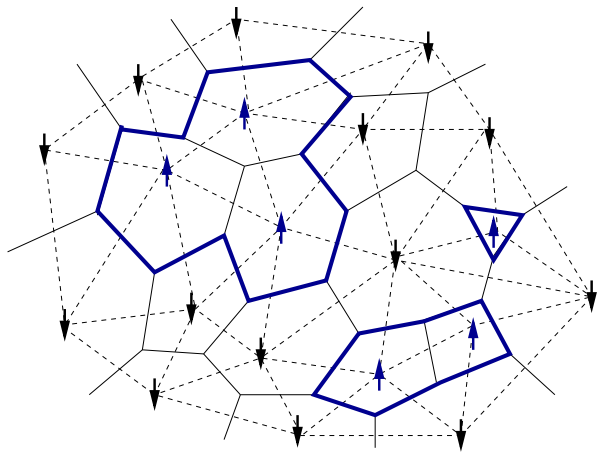}
\caption{\label{Filoops}Ising spins on the triangulated lattice, and the
corresponding domain walls drawn as loops on the dual graph.}
\end{center}
\end{figure}

Any spin configuration on the triangulation can be graphically represented as
a configuration of loops on the dual graph. Namely, for any link connecting
two spin variables of opposite sign, $\sigma_{i_*} = - \sigma_{j_*}$, one can
draw that link's dual as a part of a loop. It is easy to see that the result
will be loops surrounding domains of aligned spins (fig. \ref{Filoops}).

One can calculate the statistical weight of every loop configuration. For this
purpose, it is convenient to rewrite the partition function as~:
\begin{equation}
\Omega_{T_*}(\beta_*) = e^{3/2 N \beta_*} \sum_{\{\sigma_{i_*}\}}
e^{\beta_* \sum_{(i_*j_*)} \left( \sigma_{i_*} \sigma_{j_*} - 1 \right)} \, ,
\label{omega2}
\end{equation}
which can be obtained from (\ref{omega1}) by subtracting unity from the
link interaction energy. Since we consider triangulations without
boundaries, the numbers of links and dual links are equal, $N_L = N_{L_*}$,
and related to the number, $N$, of triangles by $N_L = 3/2 N$. The subtraction
of unity in each interaction term is compensated by adding an appropriate
constant factor in front of the sum in (\ref{omega2}). The contribution to
the sum of a term $(\sigma_{i_*} \sigma_{j_*} - 1)$ is 0 if
$\sigma_{i_*} = \sigma_{j_*}$, and 2 if $\sigma_{i_*}\ne \sigma_{j_*}$.
Thus, the sum in the exponent gives twice the number of domain wall 
links (denoted as bold links in fig. \ref{Filoops}), which
is equal to the total length $n$ of all loops on the configuration. 
Therefore~:
\begin{equation}
\Omega_{T_*}(\beta_*) = 2 \, e^{3/2 N \beta_*}
\sum_{\{L\}} e^{-2\beta_* n} \, ,
\label{omega3}
\end{equation}
where the sum runs over all loop configurations on the dual graph (which are
identical to the loop configurations of the fermionic model discussed in the
previous section). The additional factor of 2 in front of the sum reflects
the fact that each loop configuration represents two distinct spin
configurations which can be obtained from each other by a simultaneous flip of
all spins $\sigma_{i_*} \to - \sigma_{i_*}$.

For a non-spherical topology, some attention has to be paid to
non-contractible loops. Consider once more a toroidal triangulation. A
configuration with an odd number of non-contractible loops does not form a
domain wall configuration of the Ising model and therefore does not appear
in (\ref{omega3}). The same is true of the fermionic model if we perform the
GSO projection. Therefore the equivalence between the models is exact~:
\begin{equation}
{\cal Z}_T^{GSO} (K) = 2 \; e^{-3/2 N \beta_*}
\cdot \Omega_{T_*}(\beta_*) \, ,
\label{equiv1}
\end{equation}
if we set
\begin{equation}
\sqrt{3} K  = e^{-2\beta_*} \, ,
\label{KbetaS}
\end{equation}
as can be seen by comparing (\ref{Ksqrt3}) and (\ref{omega3}). This
statement holds for an arbitrary triangulation of a two-dimensional orientable
manifold without boundary.

Consider now the Ising model with spins $\sigma_i$ living on the vertices of
the dual lattice, or equivalently at the centers of the triangles of the
original manifold (in other words, the spins are located at the same spots as 
the Majorana fields $\phi_i$ discussed before). The partition function 
reads now~:
\begin{equation}
\Omega_{T}(\beta) = \sum_{\{\sigma_{i}\}}
e^{\beta \sum_{(ij)} \sigma_{i} \sigma_{j}} \, .
\label{omega0}
\end{equation}
Performing the strong coupling expansion leads to the formula~:
\begin{equation}
\Omega_{T}(\beta) = \cosh(\beta)^{3/2N\beta} \sum_{\{ \sigma_{i} \}}
\prod_{(ij)} \big( 1 + \sigma_i \sigma_j \tanh(\beta) \big) \, ,
\end{equation}
in analogy to the hopping parameter expansion (\ref{hexp}) in the Majorana
field theory. The integration rules for Ising spins~:
\begin{equation}
\frac{1}{2} \sum_{\sigma=\pm} 1 = 1 \, , \qquad
\frac{1}{2} \sum_{\sigma=\pm} \sigma = 0 \, , \qquad
\frac{1}{2} \sum_{\sigma=\pm} \sigma^2 = 1 \, ,
\end{equation} 
are completely analogous to those for the fermions (\ref{unity}),
(\ref{delta}).\footnote{One would see a difference with the fermion rules on
a lattice with vertex orders greater than three, because then one could also have
terms like $1/2 \sum_{\sigma = \pm} \sigma^4 = 1$, whereas the corresponding
terms in the Majorana model are zero, $\int {\rm d}^2 \phi \ e^{-\frac{1}{2}
\bar{\phi} \phi} \cdot \phi \phi \phi \phi = 0$. However, in our case the 
order of the dual lattice vertices is three by construction.}

Thus, calculating the strong coupling expansion, one again finds a sum over
the same loop configurations~:
\begin{equation}
\Omega_{T}(\beta) = \big(2\cosh(\beta)\big)^{3/2N\beta}
\sum_{\{ L \}'} (\tanh \beta)^{n} \, .
\label{omega4}
\end{equation}
More precisely, for a spherical lattice the loop configurations occurring in
this sum are identical to the domain wall configurations of the Ising model
defined on the triangulation. However, this is not true for topologies
of higher genus, where configurations with an odd number of non-contractible
loops from the same homotopy class occur in the strong coupling expansion
(\ref{omega4}). This is why we have put a prime on the sum, to distinguish
the set of these configurations from the set of domain walls (\ref{omega3}).
If we again take the torus as an example, we see that the sum in
(\ref{omega4}) also contains configurations with a single loop, or with an
even number of loops circling the torus in the $H_2$-direction. This kind of
loop configuration is also produced in the hopping expansion of the fermionic
model if we restrict it to the spin structure with periodic boundary
conditions. Thus, in this case we have~:
\begin{equation}
{\cal Z}_T^{(++)}(K) = \big( 2 \cosh (\beta) \big)^{-3/2 N \beta}
\cdot \Omega_{T}(\beta)
\end{equation}
if we set
\begin{equation}
\sqrt{3} K = \tanh(\beta) \, .
\label{Kbeta}
\end{equation}
The equivalence also holds for topologies of higher genus if we choose this
spin structure for the fermionic model.

As expected, both Ising models are almost dual to each other. The only
difference comes from topological contributions related to non-contractible
loops. In fact, one can make the two models exactly equivalent by a sort of a
'GSO projection' for the Ising field. Contrary to the projection in the
fermionic model, which appears as a natural option because the model
has several possible spin structures, its introduction here is somewhat
artificial.

\begin{figure}
\begin{center}
\includegraphics[height=4.5cm]{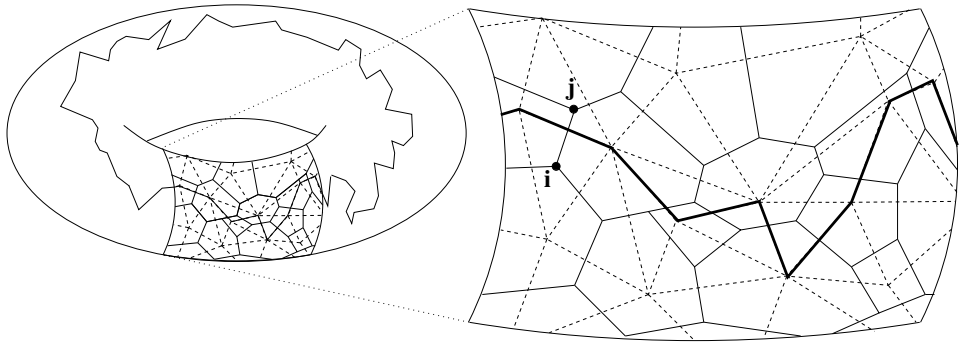}
\caption{\label{Fafline}For the dual Ising model on the torus, define an
anti-ferromagnetic line in the $H_2$-direction as a non-contractible loop
circling the torus in this direction. The Ising interaction for a given link
is defined as anti-ferromagnetic or ferromagnetic depending on whether or not
it crosses the anti-ferromagnetic line.}
\end{center}
\end{figure}

Again, take the torus as an example. Originally, we have only one version
of the Ising model, which corresponds to the spin structure $(++)$ 
(\ref{omega1}). Now, we
attempt to define a model that can reproduce the three other structures. 
Let us start with the
spin structure $(-+)$, corresponding to an anti-periodic boundary condition
in the first homotopy direction. On this lattice, choose a non-contractible
loop circling the torus once in the second homotopy direction (see
fig. \ref{Fafline}). We call this an anti-ferromagnetic line. All links
$(ij)$ that cross this line will be called anti-ferromagnetic and denoted by
$(ij)_-$. All other links will be called ferromagnetic and denoted by
$(ij)_+$. We define the partition function as follows~:
\begin{equation}
\Omega_{T}^{(-+)}(\beta) = \sum_{\{\sigma_{i}\}} 
e^{\beta \left( \sum_{(ij)_+} \sigma_{i} \sigma_{j}
- \sum_{(ij)_-} \sigma_{i} \sigma_{j} \right)} \, .
\end{equation}
In the strong coupling expansion, each ferromagnetic link contributes a
factor $+\tanh(\beta)$, and each anti-ferromagnetic link, a factor
$-\tanh(\beta)$. Each non-contractible loop in the $H_1$-direction has an
odd number of anti-ferromagnetic links, so its contribution will be
$-\tanh^n(\beta)$, whereas each contractible loop and each non-contractible
loop in the $H_2$-direction has an even number of anti-ferromagnetic links,
thus contributing $+\tanh^n(\beta)$. In other words, this prescription gives
exactly the same sign factors as those occurring for fermionic loops on the torus
with spin structure $(-+)$. In the same manner, one can also introduce an
anti-ferromagnetic line in the $H_1$-direction, to produce a model
corresponding to a $(+-)$ spin structure. Finally, a model with an
anti-ferromagnetic line in both directions gives us a $(--)$ spin structure.

Summing over all four cases, one obtains a model with a partition
function~:
\begin{equation}
\Omega_T^{GSO}(\beta) = \frac{1}{4} \left( \Omega_T^{(++)}(\beta) + 
\Omega_T^{(+-)}(\beta) + \Omega_T^{(-+)}(\beta) + \Omega_T^{(--)}(\beta) 
\right) \, ,
\end{equation}
which is exactly dual to the Ising model 
$\Omega_{T_*}^{(++)}(\beta_*)$ that has its spin variables defined
on the vertices of the triangulation (\ref{omega1}), and is equivalent to the
model of Majorana fermions with $GSO$-projection.

For a lattice size going towards infinity, the difference between
$\Omega_T^{++}(\beta)$  and $\Omega_T^{GSO}(\beta)$ becomes negligible. As
already explained, the difference comes only from the non-contractible loops. These
loops can be regarded as having a one-dimensional entropy, in the sense that
they can be ordered by a one-dimensional index that represents their position
on the lattice. Because of this, they become less and less important when the
system size grows. Therefore, in the thermodynamic limit one expects 
an exact duality between (\ref{omega1}) and (\ref{omega0}) even without
extending the model to the spin structures $(+-)$, $(-+)$, and $(--)$.
We introduced this extension here to ensure exact duality,
$i.\,e.$ a one-to-one map, between the two models even for systems of finite
size.

Generalization of this Ising model '$GSO$ construction' 
 to higher genus topologies is straightforward. In order to 
simulate a spin structure with an
antiperiodic boundary in a given direction $H_i$ (see fig. \ref{FHs}), one simply
introduces an anti-ferromagnetic line in the direction $H_j$ that crosses
$H_i$. Altogether, this creates $2^{2g}$ different spin structures.

\section{Two examples\label{sec2E}}

\begin{figure}
\begin{center}
\psfrag{e1}{{\small $X$}}
\psfrag{e2}{{\small $Y$}}
\psfrag{A}{{\footnotesize $A$}}
\psfrag{B}{{\footnotesize $B$}}
\includegraphics[height=4.5cm]{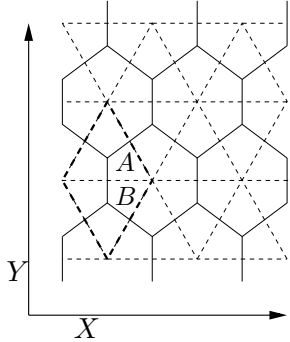}
\caption{\label{Fhoney}Fermions on the honeycomb lattice.}
\end{center}
\end{figure}

As a first example, consider the Dirac-Wilson action on a regular
triangulation of the two-dimensional plane (fig. \ref{Fhoney}). The fermions
live at the triangle centers, on a regular hexagonal
lattice. Because the lattice is flat, we can choose a global frame, $i.\,e.$
with the same directions $e_1$ and $e_2$ at each vertex. To fix the
signs, we also choose the flag assignments, which can likewise be done in a
translationally invariant way.

One can easily write down the fermionic action for this model.
Choose an elementary cell as in fig.\ref{Fhoney}. It consists
of two distinct sites~: $A$ and $B$. The lattice can be constructed
by shifting the elementary cell by multiples $i_1 d_1 + i_2 d_2$
of the fundamental shift vectors $d_1 = n_0 + n_1$, $d_2 = n_0 +n_2$
constructed from the the link vectors~: 
\begin{equation}
n_0 = (\textstyle{ 0, 1}), \quad
n_1 = \left( \textstyle{\frac{\sqrt{3}}{2}, \frac{1}{2}} \right) ,
\quad {\rm and} \quad
n_2 = \left( \textstyle{-\frac{\sqrt{3}}{2}, \frac{1}{2}} \right)\ .
\end{equation}
The components of the vectors $n_1$ and $n_2$ are expressed 
in the global frame $(X,Y)$ shown in the figure.
The position of the cell is
referred to by the double integer index $i=(i_1,i_2)$.
With this notation the action is written as~:
\begin{eqnarray}
S & = & -\frac{K}{2} \sum_{i} \sum_{d = 1}^2 \left[
\bar{\psi}_{i+d, A} (1 + n_d \cdot \gamma)
\psi_{i, B} +
\bar{\psi}_{i, B} (1 - n_d \cdot \gamma)
\psi_{i + d, A} \right] \nonumber \\
& & -\frac{K}{2} \sum_{i} \left[
\bar{\psi}_{i, A} (1 - n_0 \cdot \gamma)
\psi_{i, B} +
\bar{\psi}_{i, B} (1 + n_0 \cdot \gamma)
\psi_{i, A} \right] \nonumber \\
& & + \frac{1}{2} \sum_{i} \left[
\bar{\psi}_{i, A} \psi_{i, A} +
\bar{\psi}_{i, B} \psi_{i,B} \right] \, .
\end{eqnarray}
Since the plane is non-compact, topological effects are not relevant. From the
discussion in the previous sections we know that for Majorana fermions the
model with this action is equivalent to the Ising model with spin variables
living at the vertices and at temperature $\beta_*$ given by (\ref{KbetaS}),
and likewise to the Ising model with spins at the centers of the
triangles and at temperature $\beta$ given by (\ref{Kbeta}). The critical
temperature corresponds to the critical hopping parameter, for which the
fermions become massless. This critical value is easily found to be~:
\begin{equation}
K_{cr} = \frac{1}{3} \, ,
\end{equation}
because each vertex on the dual lattice, where the
fermions are living, has three neighbors. Thus, the critical temperatures
for the Ising models is~:
\begin{equation}
\beta_{*cr} = -\frac{1}{2} \ln \frac{\sqrt{3}}{3} \, , \quad
\beta_{cr} = \frac{1}{2} \ln (\sqrt{3}+2) \, ,
\end{equation}
in agreement with the known results \cite{b} .

A second example we want to discuss shortly here is the discretization of
the Majorana field coupled to two-dimensional gravity. It is well-known that
the integration measure over the metric field on a two-dimensional manifold
can be represented by a sum over all equilateral triangulations. If we dress
each triangulation in this sum with the fermion field, we effectively obtain 
a theory of Majorana fermions coupled to two-dimensional gravity. This theory
is given by the partition function~:
\begin{equation}
{\cal Z} (K) = \sum_T {\cal Z}_T (K) \, ,
\end{equation}
with the sum running over all triangulations with a fixed topology. For
non-spherical lattices, one should sum in addition over spin structures. 

We can use now the equivalence between the Majorana-Dirac-Wilson action and
the Ising model to substitute, triangulation by triangulation, all terms
${\cal Z}_T$ in the sum. We again obtain an exact map between the Ising model
and the model of fermions coupled to gravity. The Ising model, however, is
exactly solvable \cite{k}; in particular, 
the critical temperature is \cite{bj}~:
\begin{equation}
\beta_{cr}= \frac{1}{2} \ln \frac{108}{23} \, , \quad
\beta_{*cr}= \frac{1}{2} \ln \frac{131}{85} \, , 
\end{equation}
which means that the Majorana fermions 
are massless when the hopping parameter $K$ is~:
\begin{equation}
K_{cr} = \frac{1}{\sqrt{3}} e^{-2\beta_{*cr}} = \frac{85\sqrt{3}}{393} \, .
\label{kcr}
\end{equation}
This, again, is an exact result. The equivalence of the two models opens 
the possibility of studying numerically the properties of the Dirac-Wilson
operator coupled to gravity. In fact, one can use the Ising model as a
generator for triangulations, and then dress the configurations with local
frames and $z$ and $s$ flags to calculate ${\cal D}$ on each of them.
Since the Dirac-Wilson operator depends on the triangulation, one
gets a model of dynamical fermions interacting with the fluctuating
geometry. Using the Ising model as a Monte Carlo generator
for configurations is many orders of magnitude more efficient than a generator
referring directly to the fermionic action, since using the latter 
requires calculating the Pfaffian (\ref{Pfaffian}) in each single Monte
Carlo step, an extremely costly operation in terms of $CPU$ time.
\begin{figure}
\begin{center}
\psfrag{xx}{$\beta$}
\psfrag{yy}{$e_*$}
\includegraphics[height=6cm]{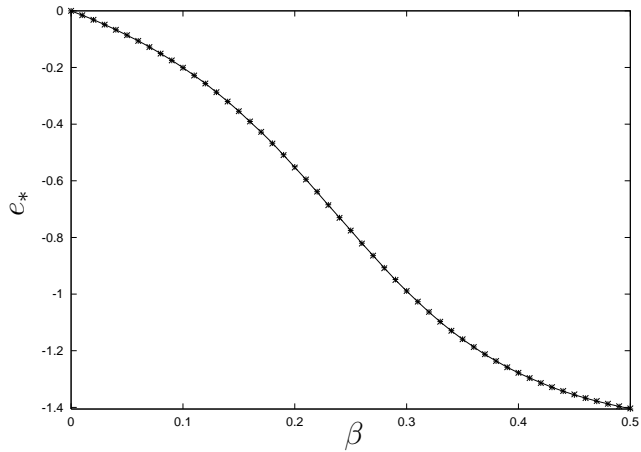}
\caption{\label{Ene} Comparison of the results for the energy
density of Ising field computed from MC simulations
of the Ising model (line) and of the corresponding qunatity 
eq.(\ref{ef}) from MC simulations of the fermionic model (crosses).
The error bars are smaller than the symbols used.}
\end{center}
\end{figure}
One can easily convince oneself by simulating small systems that the two
generators do indeed produce the same results but differ enormously in
algorithm efficiency. In fig. \ref{Ene}
we compare the average energy of the Ising
field calculated in the two different ways~:
(a) directly using the Ising model~:
\begin{equation}
e_* =
- \frac{1}{N} \left\langle
\sum_{(i_*j_*)} \sigma_{i_*} \sigma_{j_*} \right\rangle_{T} = -
\frac{1}{N} \frac{\partial }{\partial \beta_*} \ln \Omega_{T_*}
\end{equation}
or (b) using the equivalence (\ref{equiv1},\ref{KbetaS})~:
\begin{equation}
e_* = -\frac{3}{2} - \frac{1}{N} \frac{\partial K}{\partial \beta}
\, \frac{\partial\ln {\cal Z} }{\partial K} \  = \
-\frac{1}{2} -
\bigg\langle  \frac{1}{2N} \sum_a \lambda_a^{-1} \bigg\rangle_T
\label{ef}
\end{equation}
where $\lambda_a$ are eigenvalues of the Dirac--Wilson operator
${\cal D}$ on the given triangulation $T$.
In the derivation of the last formula we made use of
the relations~: \begin{equation}
\frac{\partial}{\partial K} \ln {\cal Z} =
\left\langle \frac{\partial|{\cal D}|^{1/2}}{\partial K}
\right\rangle_T =
\frac{1}{2}
\left\langle {\rm Tr} \frac{\partial D}{\partial K} {\cal D}^{-1}
\right\rangle_T
\end{equation}
The two methods yield the same results. Using the trick
with the Ising model as a generator of triangulations
one can extend the MC simulations to larger systems in order
to investigate the properties of the spectrum of the Dirac-Wilson
operator on dynamical triangulations. The results of these
investigations has been presented elsewhere \cite{bbpp}. 
Here let us
only quote a result for the finite size scaling of the
pseudocritical hopping parameter $K_*$ defined as the value
of the hopping parameter for which a mass gap is minimal.
By the mass gap we mean the center of mass
of the distribution of the smallest positive eigenvalue
of the Majorana-Dirac-Wilson operator
$\varepsilon {\cal D}$. The numerical
results can be well fitted to the finite size scaling formula~:
\begin{equation}
K_* = K_\infty + \frac{a}{N^\kappa}
\end{equation}
where $K_\infty = 0.3756(16)$, and $\kappa=1.03(30)$, $a=-0.9(5)$.
The parameter $K_\infty$ corresponds to the critical value
of the hopping parameter in the thermodynamic limit. As one can
see it agrees with the theoretical prediction
$K_{cr}=0.3746...$ given by the equation (\ref{kcr}).

\section{Conclusion\label{secD}}

The topological properties of a fermion field on discretized
two-dimensional compact manifolds were discussed at length. 
The exact equivalence
between the model of Majorana-Wilson fermions and the 
Ising model was established.
An exact duality relation for the Ising model on a compact manifold
was also found.

It would be important to generalize the construction to 
higher--di\-men\-sio\-nal
simplicial manifolds. Having done this, one would then be able to attack the
problem of quantum gravity interacting with a fermionic field. So far, it
has only been possible to couple integer spin fields to four-dimensional
simplicial gravity \cite{d2,bbkptt}. Such a theory is known to have problems 
with the continuum limit \cite{bbkp}, which could reflect the fact
that higher-dimensional gravity does not exist without a proper cocktail of
matter fields coupled to it. If this were true, the addition of fermions might
perhaps help solving these problems.

It is straightforward to generalize parts of the construction 
presented in this paper to higher dimensions. In particular, one can
associate with each four-dimensional simplex an orthonormal oriented frame
and basic rotations, and out of them one can easily build the transition
matrices and spin connections. However, the problem of lifting this
construction to the half-integer representation leads to additional 
complications.

One of the reasons is that the topological problem is by itself
more complicated in four dimensions. The question of whether a 
manifold admits a spin structure, which is equivalent to the 
question of whether it is possible to define globally a
Dirac operator on it, is in general related to the existence of a
non-trivial second Stiffel-Witney form \cite{top1,top2}. 
For two- and three-dimensional manifolds, this reduces to 
the orientability question. In four dimensions, however, 
there are manifolds, like for example the projective
space $CP (C^2)$, which are orientable but possess a non-trivial
Stiffel-Witney form, and thus do not admit any spin structure. 
In an attempt of extending our construction to a higher dimensional 
manifold not admitting any spin structure, the topological 
obstruction would manifest itself as the impossibility to 
adjust the local degrees of freedom so as to assign positive 
signs to all elementary plaquettes.

\section*{Acknowledgments}
We thank Joachim Tabaczek for many discussions. 
This work was supported in part by the EC IHP 
grant HPRN-CT-1999-00161 and
by the Polish Government Project (KBN)  2P03B 01917.

\end{document}